\newcommand{\review}{\iffalse} %Final  (two column)

\review
\documentclass[preprint,review,12pt]{elsarticle}
\else
\documentclass[final,5p,twocolumn,times]{elsarticle}
\fi

\usepackage{graphicx} % Include figure files
\usepackage{amsmath}
\usepackage{amssymb}
\usepackage{esint}
\usepackage{textcomp}
\usepackage{xcolor}
\definecolor{myblue}{rgb}{0.0, 0.0, 0.6}
\usepackage{hyperref}
\hypersetup{
  colorlinks = true,
  citecolor  = myblue,
  linkcolor  = myblue,
  urlcolor   = myblue
}

\usepackage{lineno}
\usepackage{setspace}

\journal{Nuclear Instruments and Methods in Research A}

\begin{document}
\review \linenumbers \fi
%\linenumbers

\begin{frontmatter}  
\title{ Evaluation of the beam-induced depolarization of the HJET target at the EIC
  %\tnoteref{t1}
}%
%\tnotetext[t1]{This manuscript has been authored by employees of Brookhaven Science Associates, LLC under Contract No. DE-SC0012704 with the U.S. Department of Energy}

\author{A.~A.~Poblaguev}\corref{cor1}\ead{poblaguev@bnl.gov}\cortext[cor1]{Corresponding author}

\address{Brookhaven National Laboratory, Upton, New York 11973, USA}

\date{\today}% It is always \today, today,

\begin{abstract}
The Polarized Atomic Hydrogen Gas Jet Target (HJET) has played a central role in the absolute calibration of proton beam polarization at RHIC and is foreseen as a key component of the hadron polarimetry program at the future Electron--Ion Collider (EIC). The substantially higher beam current, reduced bunch spacing, and shorter bunch length planned for EIC operation motivate a careful reassessment of possible beam-induced depolarization of the jet target. 
In this paper, the depolarization of ground-state hydrogen atoms caused by the time-dependent magnetic field of the circulating polarized proton beam is quantitatively evaluated. The hydrogen atom is treated as a four-level hyperfine system in a holding magnetic field, and transitions driven by harmonic components of the bunch-induced magnetic field are analyzed using time-dependent quantum-mechanical evolution along atomic trajectories. Numerical tracking of hydrogen atoms through the beam region is performed using nominal EIC beam parameters.
It is shown that, for a holding field of $120\,\mathrm{mT}$ (as used at RHIC), the resulting depolarization of the jet target at the EIC is negligibly small, $\lesssim 0.01\%$, and well below the level relevant for EIC polarization accuracy requirements. The stability of this result with respect to plausible variations of the EIC proton beam parameters is also evaluated. In addition, possible effects under alternative experimental conditions are also examined.
\end{abstract}

\begin{keyword}
Electron--Ion Collider \sep
Hadron polarimetry \sep
Polarized hydrogen target \sep
Beam-induced depolarization \sep
Breit--Rabi polarimeter 
\end{keyword}

\end{frontmatter}

\section{Introduction}

High-energy $41$--$275\,\mathrm{GeV}$ polarized ($\gtrsim70\%$) proton beams are planned for the Electron--Ion Collider (EIC) \cite{Accardi:2012qut,Willeke:2021ymc}. The Science Requirements for hadron polarimetry at the EIC \cite{AbdulKhalek:2021gbh} include a precise calibration of the measured proton beam polarization, with systematic uncertainties better than
\begin{equation}
  \sigma_P^\textrm{syst}/P \lesssim 1\%.
  \label{eq:EICreq}
\end{equation}

Since hadron polarimetry at the Relativistic Heavy-Ion Collider (RHIC) has been successfully performed for more than two decades, it is considered a natural starting point for the EIC hadron polarimetry program.

Within the RHIC Spin Program \cite{Bunce:2000uv,Aschenauer:2015eha}, several polarimeters are employed, including the 200\,MeV absolute proton--carbon polarimeter at the LINAC \cite{Zelenski:2013zxa,Poblaguev:2025ZO}, the relative proton--carbon recoil polarimeters \cite{Huang:2006cs} at the AGS and RHIC, and the Polarized Atomic Hydrogen Gas Jet Target (HJET) \cite{Zelenski:2005mz,Poblaguev:2020qbw}. These instruments are used to monitor the beam polarization and to provide essential information to the RHIC experiments, such as the polarization value, polarization profile, proton spin tilt, and polarization decay time. The primary role of the HJET is the precise absolute calibration of the proton beam polarization.

During polarized proton runs at RHIC, the systematic uncertainty of the measured proton beam polarization was evaluated to be $\sigma_P^\text{syst}/P \sim 0.5\%$ \cite{Poblaguev:2019saw,Poblaguev:2020qbw}, which satisfies the EIC requirement given by Eq.~\eqref{eq:EICreq}.

\begin{table*}[htbp]
	\centering
	\caption{\label{tab:BeamPar}
	Beam parameters essential for the evaluation of beam-induced depolarization for RHIC flattop and for EIC injection and flattop nominal conditions \cite{Rathmann:2025jgp}. For both machines, the circumference is $3833.85\,\mathrm{m}$. IP\,12 and IP\,4 denote the interaction points where the HJET is located at RHIC and will be located at the EIC, respectively. For simplicity, the transverse beam profile at the EIC is approximated in this paper by a round distribution with rms radius $\sigma_r=(\sigma_x\sigma_y)^{1/2}$.}
	\begin{tabular}{llc|r|rr}
		\multicolumn{3}{c|}{}  & RHIC at IP\,12 & \multicolumn{2}{c}{EIC at IP\,4} \\
		Parameter & Notation & Unit  & flattop & injection & flattop \\
		\hline
		Beam energy             & $E_{\text{beam}}$ & GeV      & 255     &  23.5  &  275     \\
		Number of bunches       & $N_\text{b}$      & --       & 120     & 290    & 1160     \\
		Protons per bunch       & $N_p$             & $10^{10}$ & 20     & 27.6   & 6.9      \\
		Temporal bunch length   & $\sigma_t$        & ns       & 1.835   & 0.801  & 0.200    \\
		Bunch spacing           & $\tau_\text{b}$   & ns       & 106.598 & 44.144 & 11.027   \\
		Bunch frequency         & $f_\text{b}$      & MHz      & 9.381   & 22.653 & 90.683   \\
		Average beam current    & $I_{\text{avg}}$  & A        & 0.301   & 1.002  & 1.003    \\
		Transverse beam size    & $\sigma_x$        & mm       & --      & 3.513  & 1.610    \\
		                        & $\sigma_y$        & mm       & --      & 0.689  & 0.268    \\
		Radial beam size        & $\sigma_r$        & mm       & 0.23    & 1.566  & 0.656    \\
	\end{tabular}
\end{table*}

As shown in Table~\ref{tab:BeamPar}, at the EIC the beam current will be increased by more than a factor of three compared to RHIC, while the bunch spacing and the bunch length will be reduced by nearly an order of magnitude. Consequently, a reassessment of the HJET performance under the new operating conditions is required.

Possible effects of the short bunch spacing on proton beam polarization measurements with the HJET recoil spectrometer were previously evaluated in Refs.~\cite{Poblaguev:2020duy,Poblaguev:2024yhl} using an emulation of the EIC bunch structure based on HJET experimental data obtained at RHIC. In these studies, however, the jet target polarization --- which is determined, under certain assumptions, by the HJET Breit--Rabi polarimeter (BRP) --- was assumed to be well known.

Recently, beam-induced depolarization \cite{HERMES:1998twm} of the jet target itself was evaluated in Ref.\,\cite{Rathmann:2025jgp}, where it was claimed that unavoidably large depolarization would occur at the EIC if a 120\,mT holding field (the value used at RHIC) is applied. Since many questions have arisen regarding the methodology used in \cite{Rathmann:2025jgp} and, consequently, the reliability of this result, an alternative analysis, based on the numerical solution of the fundamental quantum-mechanical equations governing the depolarization process, is presented here. For a direct comparison of the two approaches, this paper adopts the same beam parameters (given in Table~\ref{tab:BeamPar}) as in Ref.\,\cite{Rathmann:2025jgp}.
   
  In such a study, it is essential to compare the depolarization effects relevant for the recoil spectrometer with those affecting the BRP. Therefore, the paper also includes a discussion of basic HJET design-related systematic effects relevant to these measurements.

As a prerequisite, a brief description of the hyperfine structur<e of a ground-state hydrogen atom in the holding-field magnet is given in Section~\ref{sec:hfs}. The RHIC HJET design and performance, including a discussion of the optimization of the BRP measurement algorithm, are reviewed in Section~\ref{sec:HJET}. The methodology used to evaluate beam-induced depolarization is described in Section~\ref{sec:Meth}. In Section~\ref{sec:Est}, it is shown that the expected depolarization of the HJET target at the EIC is negligibly small. The stability of this result with respect to plausible variations of the proton beam parameters is discussed in Section~\ref{sec:Stability}. The method used is benchmarked in Section~\ref{sec:Benchmark}.

\section{The hyperfine structure of the ground-state hydrogen\label{sec:hfs}}

The hyperfine structure of the hydrogen atom ground state in an external magnetic field $\boldsymbol{B}$ is described by the electron-- and proton--spin-dependent Hamiltonian, which can be written in terms of Pauli matrices as \cite{Feynman2011}
\begin{equation}
  H_0 = E_0 + \frac{E_\text{hfs}}{4}\,\boldsymbol{\sigma}^e\!\cdot\!\boldsymbol{\sigma}^p
  -\left(\mu_e\boldsymbol{\sigma}^e + \mu_p\boldsymbol{\sigma}^p\right)\!\cdot\!\boldsymbol{B},
  \label{eq:H0}
\end{equation}
where $\mu_e$ and $\mu_p$ are the magnetic moments of the electron and proton, respectively. Numerically \cite{Mohr:2024kco},
\begin{equation}
  \frac{\mu_e}{2\pi\hbar} = -14.012\,\text{MHz/mT},\qquad
  \frac{\mu_p}{2\pi\hbar} = 0.021\,\text{MHz/mT}.
\end{equation}
In the following, the proton magnetic moment is neglected.

\begin{figure}[t]
\begin{center}
\includegraphics[width=0.95\columnwidth]{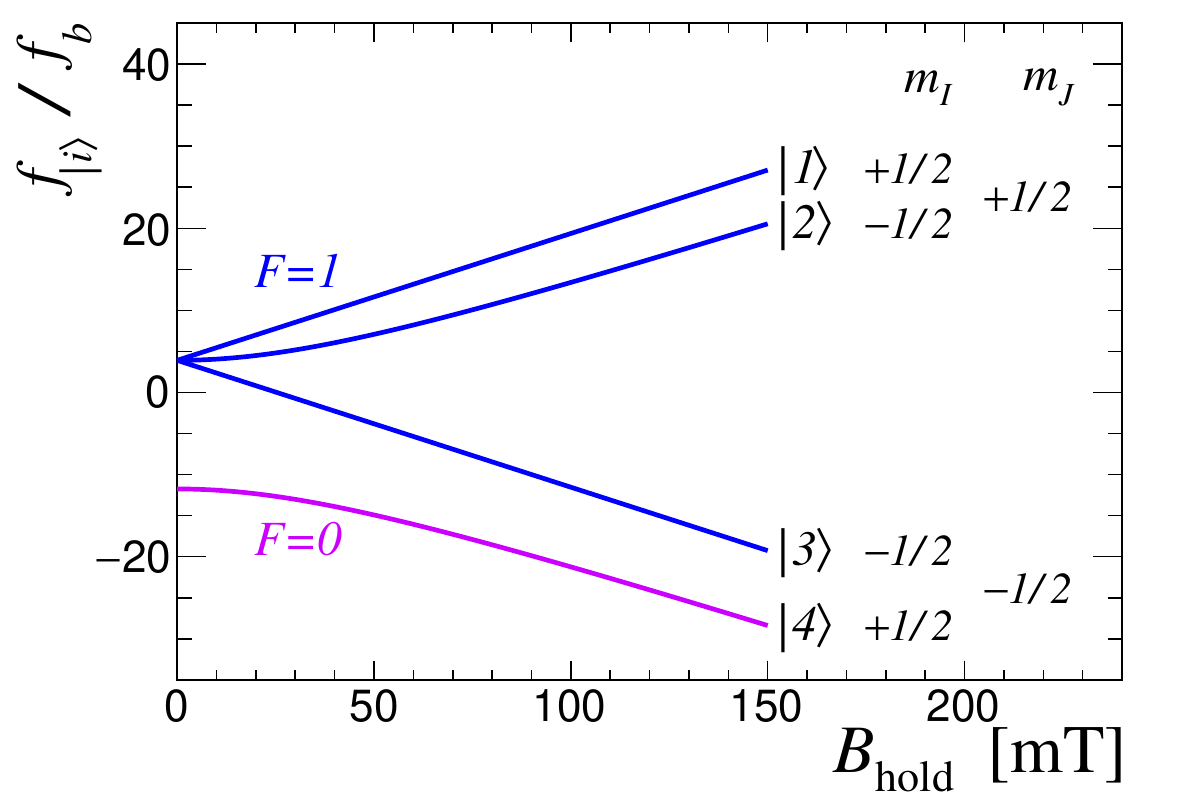}
\end{center}
\caption{\label{fig:HFS}
Breit--Rabi diagram for ground-state hydrogen ($1S_{1/2}$). The projection of the total spin $F$ onto the holding magnetic field $B_\text{hold}$ is given by $m_F=m_I+m_J$, where $m_I$ and $m_J$ are the proton and electron spin projections, respectively, in the high-field limit ($\sin\theta\to0$ in Eq.\,\eqref{eq:Hstates}).
}
\end{figure}

Due to the electromagnetic multipole interaction between the proton and the electron cloud, the ground state of the hydrogen atom is split into sublevels characterized by the total atomic angular momentum quantum number $F$: a triplet state ($F=1$) and a singlet state ($F=0$), as illustrated in Fig.\,\ref{fig:HFS}. The separation between the $F=1$ and $F=0$ levels is known with very high precision \cite{Diermaier:2016fsy},
\begin{equation}
 \frac{E_\text{hfs}}{2\pi\hbar} = f_\text{hfs}
 = 1420405748.4 \pm 3.4_\text{stat} \pm 1.6_\text{syst}\,\text{Hz}.
\end{equation}
In energy units, this corresponds to $E_\text{hfs}\approx5.9\times10^{-6}\,\text{eV}$.

In the presence of a holding magnetic field $B_\text{hold}$, the triplet state undergoes Zeeman splitting. The corresponding energy levels, expressed in frequency units, can be approximated as
\begin{equation}
  \begin{aligned}
    f_{|1\rangle}(B_\text{hold}) &= \frac{f_\text{hfs}}{2}\!\left(+\frac{1}{2} + x\right), \\
    f_{|2\rangle}(B_\text{hold}) &= \frac{f_\text{hfs}}{2}\!\left(-\frac{1}{2} + \sqrt{1+x^2}\right), \\
    f_{|3\rangle}(B_\text{hold}) &= \frac{f_\text{hfs}}{2}\!\left(+\frac{1}{2} - x\right), \\
    f_{|4\rangle}(B_\text{hold}) &= \frac{f_\text{hfs}}{2}\!\left(-\frac{1}{2} - \sqrt{1+x^2}\right),
  \end{aligned}
  \label{eq:f}
\end{equation}
with
\begin{equation}
  x = \tan^{-1}{2\theta} =  \frac{B_\text{hold}}{B_c}, \qquad
  B_c = \frac{f_\text{hfs}}{2|\mu_e/2\pi\hbar|} = 50.7\,\text{mT}.
  \label{eq:theta}
\end{equation}

The spin structure of the hyperfine eigenstates and the corresponding nuclear polarizations are given by
\begin{equation}
  \begin{aligned}
    |1\rangle &= |e^\uparrow p^\uparrow\rangle, &
    P_{|1\rangle} &= +1, \\
    |2\rangle &= \cos\theta\,|e^\uparrow p^\downarrow\rangle
    + \sin\theta\,|e^\downarrow p^\uparrow\rangle, &
    P_{|2\rangle} &= -\cos(2\theta), \\
    |3\rangle &= |e^\downarrow p^\downarrow\rangle, &
    P_{|3\rangle} &= -1, \\
    |4\rangle &= \cos\theta\,|e^\downarrow p^\uparrow\rangle
    - \sin\theta\,|e^\uparrow p^\downarrow\rangle, &
    P_{|4\rangle} &= +\cos(2\theta),
  \end{aligned}
  \label{eq:Hstates}
\end{equation}
where $\uparrow,\downarrow$ denote spin projections along the magnetic-field direction.

A weak external oscillating magnetic field $\boldsymbol{B}_\text{osc}\cos(\omega t)$ can induce resonant transitions between states $|i\rangle$ and $|j\rangle$ if
\begin{equation}
  \frac{\omega}{2\pi} = f_{ij}(B_\text{hold})
  = f_{|i\rangle}(B_\text{hold}) - f_{|j\rangle}(B_\text{hold}),
  \label{eq:f_ij}
\end{equation}
a mechanism widely used for polarizing atomic hydrogen beams \cite{Haeberli:1967zr}. Conversely, when such a polarized atomic hydrogen beam is employed as a target in the HJET polarimeter, it may be depolarized by the time-dependent magnetic field generated by the circulating proton beam bunches.

\subsection{External perturbation of a ground-state hydrogen atom}

A ground-state hydrogen atom can be treated as a four-level quantum system with eigenstates $|n\rangle$ and eigenvalues $E_{|n\rangle}$ of the time-independent Hamiltonian $H_0$ in Eq.\,\eqref{eq:H0}. In the presence of a weak time-dependent perturbation $H_1(t)$, the Schr\"odinger equation in the interaction picture leads to coupled equations for the expansion coefficients $a_n(t)$ \cite{LandauQM}:
\begin{equation}
  \left[ H_0 + H_1(t) - i\hbar\frac{d}{dt} \right]
  \sum_n a_n(t)\,|n\rangle e^{-iE_n t/\hbar} = 0.
  \label{eq:da/dt}
\end{equation}
The time evolution of the atomic state is governed by
\begin{equation}
  i\hbar\frac{d a_k(t)}{dt}
  = \sum_n {\cal M}_{kn}(t)\,a_n(t)\,e^{i\omega_{kn} t},
  \label{eq:an(t)}
\end{equation}
where ${\cal M}_{kn}(t)=\langle k|H_1(t)|n\rangle$ is the matrix element of the perturbation and $\omega_{kn}=(E_k-E_n)/\hbar$. For specified matrix elements and initial conditions $a_n(0)$, the system~\eqref{eq:an(t)} can be solved exactly, for example, by numerical integration.

If the hydrogen atom is in an oscillating magnetic field with angular frequency $\omega$, then
\begin{equation}
  {\cal M}_{kn}(t) = \hbar\omega_R\cos{\omega t},
\end{equation}
where $\omega_R$ is commonly referred to as the Rabi (angular) frequency.
  
To analyze the behavior of the system in the vicinity of a resonant transition (i.e., if $\omega\approx\omega_{ij}$ for some transition $|i\rangle\to|j\rangle$), consider a two-level subsystem $ij$ initially prepared in state $|i\rangle$, with $a_i(0)=1$ and $a_j(0)=0$. For small detuning,
$|\Delta\omega| = |\omega-\omega_{ij}| \ll |\omega_{ij}|$, and for short interaction times $t \ll |\Delta\omega|^{-1}$ (but still much larger than $|2\omega_{ij}|^{-1}$), one finds
\begin{equation}
  a_j(t) = -i\frac{\omega_R}{2}
  \int_0^t dt'a_i(t')\left(
    e^{-i\Delta\omega t'} + e^{-i2\omega_{ij}t'}
  \right)
  \approx -i\frac{\omega_R t}{2}.
  \label{eq:coherence}
\end{equation}
For estimate, it was replaced in the integrand: $a_i(t^\prime)\to1$ (upper limit for the absolute value), $\exp{(-i\Delta\omega t^\prime)}\to1$ (since $\Delta\omega t^\prime\ll1$), and $\exp{(-i2\omega_{ij}t^\prime)}\to0$ (very frequently oscillating term).

If total time  the atom is under the resonant condition is $t_\text{int}$, then
\begin{equation}
  \varepsilon = \omega_R t_\text{int}/2
\end{equation}
can serve as a parametrization of maximally possible resonant $|i\rangle\to|j\rangle$ transition probability. In particular, if $\varepsilon^2<10^{-3}$, which corresponds to the transition probability of less than 0.1\%, the resonant transition $|i\rangle\to|j\rangle$ can be disregarded (in a study of the beam induced depolarization at HJET) and, consequently, one can approximate ${\cal M}_{ij}(t)\to0$. Since $a_i(t^\prime)\to1$ was used in Eq.\,\eqref{eq:coherence}, the conclusion can be straigtforwardly expanded to a more general case of the multi-level system \eqref{eq:an(t)}.

For longer interaction times $t\gtrsim\Omega^{-1}$ in Eq.\,\eqref{eq:coherence}, both transitions
$|i\rangle \leftrightarrow |j\rangle$ must be treated consistently, which leads to the well-known result~\cite{Rabi:1937dgo}
\begin{equation}
  a_j(t) = -i\,\frac{\omega_R}{\Omega}
  \sin\!\left(\frac{\Omega t}{2}\right), 
  \qquad
  \Omega = \sqrt{\omega_R^2 + \Delta\omega^2}.
  \label{eq:Prob}
\end{equation}
This expression is fully consistent with Eq.~\eqref{eq:coherence} in the limit of short times $t \ll 1/\Omega$.

In the limit $\omega_R\ll\omega$ and $|\Delta\omega|\ll\omega$, it is convenient to work in a rotating reference frame, defined by $a_i(t)=\alpha_i(t)e^{-i\omega t}$ and $a_f(t)=\alpha_f(t)e^{i\omega t}$. The two-level system then obeys
\begin{equation}
  \begin{aligned}
    i\,\frac{d\alpha_i(t)}{dt} &=
    -\frac{\Delta\omega}{2}\,\alpha_i(t)
    + \frac{\omega_R}{2}\,\alpha_f(t), \\
    i\,\frac{d\alpha_f(t)}{dt} &=
    \frac{\omega_R}{2}\,\alpha_i(t)
    + \frac{\Delta\omega}{2}\,\alpha_f(t).
  \end{aligned}
  \label{eq:alpha}
\end{equation}

For $\alpha_i(0)=1$ and $\alpha_f(0)=0$, and \emph{time-independent} $\omega_R$ and $\Delta\omega$, the solution can be readily derived:
\begin{align}
  \alpha_i(t) &=
  \cos\!\left(\frac{\Omega t}{2}\right)
  + i\,\frac{\Delta\omega}{\Omega}
  \sin\!\left(\frac{\Omega t}{2}\right),
  \label{eq:alpha+} \\
  \alpha_f(t) &=
  -i\,\frac{\omega_R}{\Omega}
  \sin\!\left(\frac{\Omega t}{2}\right),
  \label{eq:alpha-}
\end{align}
in full consistency with Eq.\,\eqref{eq:Prob}.

For a hydrogen atom traversing the proton beam in the HJET, both $\omega_R(t)$ and $\Delta\omega(t)$ are, in general, time-dependent. Consequently, Eq.\,\eqref{eq:Prob} can be used only for preliminary estimates, while the full time-dependent problem must be addressed by numerical integration.

\section{The RHIC HJET\label{sec:HJET}}

\begin{figure}[t]
  \begin{center}
    \review
    \includegraphics[width=0.6\columnwidth]{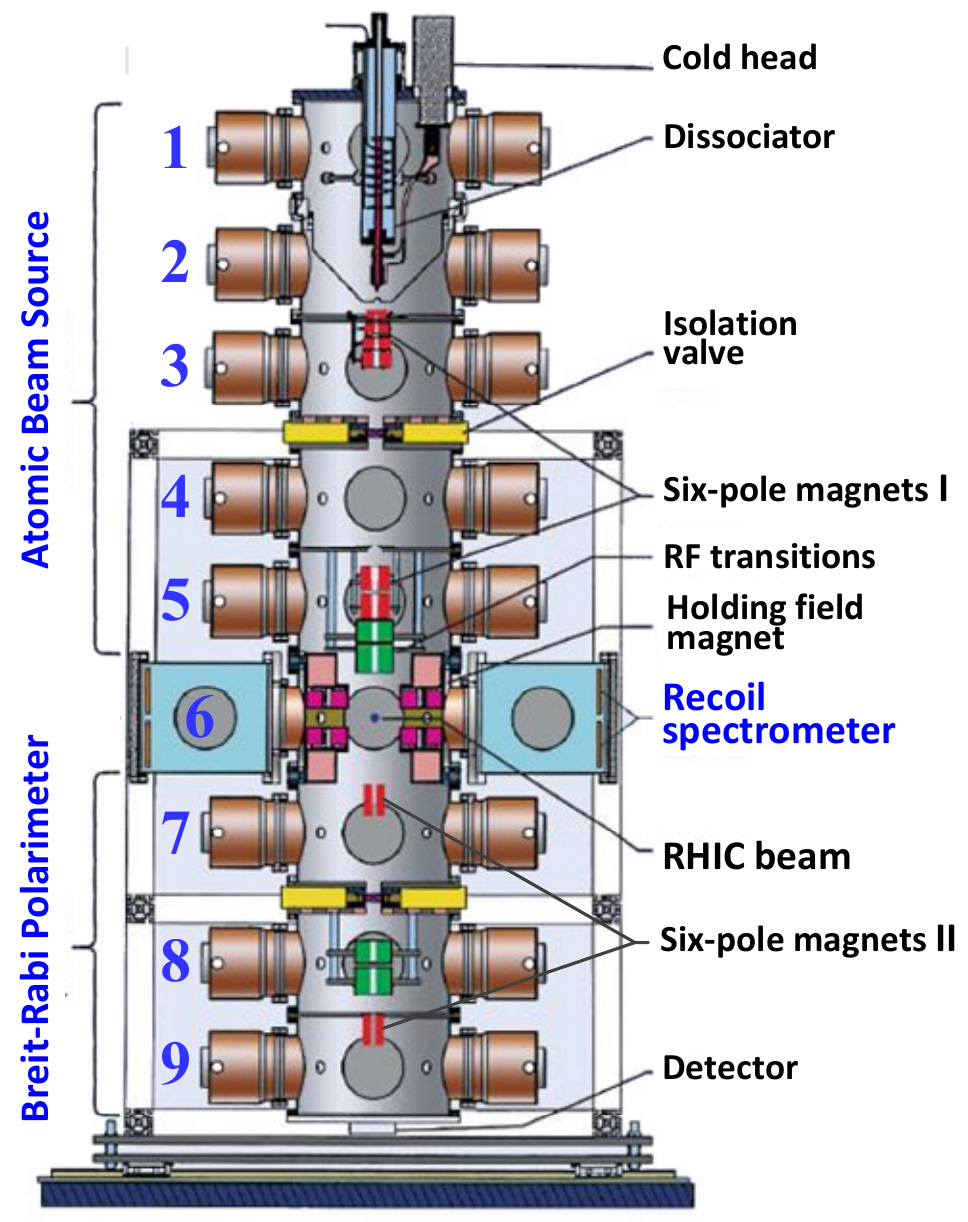}
    \else
    \includegraphics[width=0.8\columnwidth]{Fig02_ABS.pdf}
    \fi
  \end{center}
\caption{Beam view of the Polarized Atomic Hydrogen Gas Jet target polarimeter at RHIC.}
\label{fig:HJET}
\end{figure}

The RHIC HJET polarimeter \cite{Zelenski:2005mz,Wise:2006xj} consists of nine vertically arranged pumping stages (chambers), as depicted in Fig.~\ref{fig:HJET}. The three main components of the polarimeter are the Atomic Beam Source (ABS, chambers~1--5), the recoil spectrometer (chamber~6), and the Breit--Rabi polarimeter (BRP, chambers~7--9). A powerful vacuum system, combined with a carefully designed system of separating magnets, enables delivery of an atomic beam with an intensity of
$1.2\times10^{17}\,\text{atoms/s}$. The atomic beam is focused at the collision point, where the jet thickness is approximately
$1.2\times10^{12}\,\text{atoms/cm}^2$, and the transverse profile can be approximated by a Gaussian distribution with
$\sigma\approx2.6\,\text{mm}$.

\subsection{The atomic beam source}

The H$_2$ gas injected into the ABS is dissociated into four nearly equally populated hyperfine states of hydrogen atoms,
\begin{flalign}
 \text{Dissociator:}\qquad\bigg\{n_1,~~n_2,~~n_3,~~n_4\bigg\}.&&
\end{flalign}
The atoms expand through a cold nozzle and enter the sextupole separating magnet system. Based on the Stern--Gerlach method, the sextupole magnets (with a typical field strength of 1.6--1.7\,T) separate atoms with electron spin $+1/2$ (hyperfine states $|1\rangle$ and $|2\rangle$) from those with spin $-1/2$ ($|3\rangle$ and $|4\rangle$). In addition, the sextupole system focuses the electron spin $+1/2$ atoms into the target region.

Omitting an overall normalization factor, the atomic-beam hyperfine-state populations at the entrance to the RF transition unit can be approximated as
\begin{flalign}
  \text{Sep.\,Magnet~I:}\qquad\bigg\{1-\nu,~~1+\nu,~~\epsilon_\text{I},~~\epsilon_\text{I}\bigg\}.&&
  \label{eq:SepMagI}
\end{flalign}
Here, the parameter $\nu$ accounts for a small difference in the transmission efficiencies of states $|1\rangle$ and $|2\rangle$ through the separating magnet system. In the HJET BRP software, this asymmetry is hard-coded as
\begin{equation}
   \frac{n_2}{n_1}=1.00239 \quad \Rightarrow \quad \nu\approx0.12\%.
\end{equation}
For comparison, in the HERMES polarized hydrogen target \cite{HERMESTargetGroup:2001qci}, this ratio was experimentally determined to be
$n_2/n_1 = 1.029 \pm 0.0015$ ($\nu \approx 1.5\%$). As will be shown below, for $|\nu| < 2.5\%$, neglecting $\nu$ in the HJET data analysis has a negligible impact ($<0.1\%$) on the beam polarization measurement.

The parameter $\epsilon_\text{I}$ in Eq.~\eqref{eq:SepMagI} represents the residual transmission of states $|3\rangle$ and $|4\rangle$; $\epsilon_\text{I}=0$ corresponds to an ideal sextupole separating magnet system. For the HJET ABS design, it was estimated~\cite{Wise:2006xj} that $\epsilon_\text{I}<0.002$. Based on the calculations reported in Ref.\,\cite{Mikirtychyants:2012nh}, one finds for the ANKE experiment ABS: $\epsilon_\text{I}=0.0025$ for the state $|3\rangle$ and $\epsilon_\text{I}=0.001$ for the state $|4\rangle$. Owing to the smallness of $\epsilon_\text{I}$, no distinction is made here between the efficiencies for the $|3\rangle$ and $|4\rangle$ states.

If only the Strong Field Transition (SFT), which provides an adiabatic transition between states $|2\rangle$ and $|4\rangle$, is enabled (SFT=On, WFT=Off), the hyperfine-state populations become
\begin{flalign}
 \text{SFT:}\quad\bigg\{1-\nu,\ (1+\nu)\epsilon_{24}+\epsilon_\text{I},\ \epsilon_\text{I},\ (1+\nu)(1-\epsilon_{24})\bigg\},&&
\end{flalign}
where $\epsilon_{24}$ denotes the SFT transition inefficiency. Consequently, the atomic hydrogen beam acquires the polarization
\begin{equation}
  P_+ =
  \frac{(1-\epsilon_\text{I})\cos^2\theta
        - \epsilon_{24}\cos2\theta
        - \nu\left(\sin^2\theta+\epsilon_{24}\cos2\theta\right)}
       {1+\epsilon_\text{I}}.
   \label{eq:P+}
\end{equation}

For the RHIC HJET holding field of $B_\text{hold}=120\,\text{mT}$ and using Eq.\,\eqref{eq:theta}
\begin{equation}
  \cos^2\theta=0.961,\quad
  \sin^2\theta=0.039,\quad
  \cos2\theta=0.921.
\end{equation}

Similarly, for the Weak Field Transition configuration (WFT=On, SFT=Off), one obtains
\begin{flalign}
  \text{WFT:}\quad\bigg\{(1-\nu)\epsilon_{13}+\epsilon_\text{I},\ 1+\nu,\ (1-\nu)(1-\epsilon_{13}),\ \epsilon_\text{I}\bigg\} &&
\end{flalign}
and
\begin{equation}
  -P_- =
  \frac{(1-\epsilon_\text{I})\cos^2\theta
        - \epsilon_{13}
        - \nu\left(\sin^2\theta-\epsilon_{13}\right)}
       {1+\epsilon_\text{I}}.
       \label{eq:P-}
\end{equation}

At RHIC, the jet target polarization was typically reversed every 300\,s during HJET operation. Between periods with hydrogen nuclear spin oriented up and down, both RF transitions (SFT and WFT) were turned on,
\begin{flalign}
  &\text{SFT+WFT:}\quad\bigg\{(1-\nu)\epsilon_{13}+\epsilon_\text{I},\ (1+\nu)\epsilon_{24}+\epsilon_\text{I},\ &\nonumber \\
  &\qquad\qquad(1-\nu)(1-\epsilon_{13}),\ (1+\nu)(1-\epsilon_{24})\bigg\},
  \\
  &P_0=
  \frac%
 {-\sin^2\theta\,\left[(1\!+\!\nu)(1\!-\!2\epsilon_{24})\!-\!\epsilon_\text{I}\right] +\nu(1\!-\!\epsilon_{13}\!-\!\epsilon_{24})+\epsilon_{13}\!-\!\epsilon_{24}}%
 {1+\epsilon_\text{I}},
\end{flalign}
for a short 30\,s interval, allowing a BRP-based determination of corrections to the nominal jet polarizations $\pm\cos^2\theta$ defined by the holding-field value.

\subsection{The recoil spectrometer}

  In the recoil spectrometer, the vertically polarized proton beam, consisting of alternatively polarized bunches, is scattered off the jet target, and the low-energy recoil protons are detected by left/right symmetric silicon detectors located at $90^\circ$ with respect to the beam direction \cite{Poblaguev:2020qbw}. If only one proton (beam or jet) is (vertically) polarized, the numbers of events counted in the left (L) and right (R) detectors depending of the spin direction ($\uparrow\downarrow$) can be approximated as:
  \begin{equation}
    \begin{aligned}
      N_R^\uparrow   = N_0\times(1+a+a\tilde{\nu})\,(1+\lambda)\,(1+\varkappa), \\
      N_R^\downarrow = N_0\times(1-a-a\tilde{\nu})\,(1-\lambda)\,(1+\varkappa), \\
      N_L^\uparrow   = N_0\times(1-a+a\tilde{\nu})\,(1+\lambda)\,(1-\varkappa), \\
      N_L^\downarrow  = N_0\times(1+a-a\tilde{\nu})\,(1-\lambda)\,(1-\varkappa), 
    \end{aligned}
    \label{eq:NRL}
  \end{equation}
  where $N_0$ is a normalization factor, $\lambda$ is the beam intensity asymmetry, and $\varkappa$ is the detector acceptance asymmetry. The spin asymmetry is parameterized as
\begin{equation}
  a = A_\text{N}\frac{|P_+| + |P_-|}{2}, \qquad
  \tilde{\nu} = \frac{|P_+| - |P_-|}{|P_+| + |P_-|},
\end{equation}
where $A_\text{N}$ is the analyzing power. In HJET measurements, $a < 0.04$ \cite{Poblaguev:2020qbw}, and mathematically $|\tilde{\nu}| \le 1$.

If $\tilde{\nu}=0$ system \eqref{eq:NRL} can be solved exactly \cite{Poblaguev:2020qbw}, 
\begin{equation}
  a = \frac%
  {\sqrt{N_R^\uparrow N_L^\downarrow} - \sqrt{N_L^\uparrow N_R^\downarrow} }%
  {\sqrt{N_R^\uparrow N_L^\downarrow} + \sqrt{N_L^\uparrow N_R^\downarrow} },
  \label{eq:a} 
\end{equation}
and analogous expressions for $\lambda$ and $\varkappa$.

Since term $a\tilde{\nu}$ is strongly correlated with $\lambda$ in Eq.\,\eqref{eq:NRL}, formula \eqref{eq:a} provides accurate evaluation of asymmetry $a$ even if $\tilde{\nu}\ne0$:
\begin{align}
  a_\text{calc}& = \frac%
  {\sqrt{(1+a)^2-a^2\tilde{\nu}^2} - \sqrt{(1-a)^2-a^2\tilde{\nu}^2} } 
  {\sqrt{(1+a)^2-a^2\tilde{\nu}^2} + \sqrt{(1-a)^2-a^2\tilde{\nu}^2} }
  \nonumber \\ &=
  a\times\left[ 1+a^2\tilde{\nu}^2 + {\cal O}\left(a^4\tilde{\nu}^2\right) \right]
  \nonumber \\ &\approx a.
  \label{eq:ac} 
\end{align}

In the HJET measurements, both, beam and jet, asymmetries are determined concurrently. 
Since identical particles are scattered, the elastic transverse analyzing powers for the beam, $A_N^\text{beam}(t)$, and for the jet, $A_N^\text{jet}(t)$, as functions of the momentum transfer $t$, are identical. Moreover, if background events are properly eliminated from the $\mathit{pp}$ elastic-scattering data, one has $\langle{A_N^\text{beam}}\rangle\!=\!\langle{A_N^\text{jet}}\rangle$ \cite{Poblaguev:2020qbw}, because exactly the same events are used to calculate both beam and jet asymmetries. Therefore,
\begin{equation}
  P_\text{beam} = \frac{a_\text{beam}}{a_\text{jet}}\,P_\text{jet}.
  \label{eq:Pbeam}
\end{equation}
To satisfy the EIC requirement \eqref{eq:EICreq}, the jet target polarization must be known with an accuracy significantly better than 1\%.

As it was demonstrated in Eq.\,\eqref{eq:ac}, 
only spin-direction-averaged values of the beam and jet polarizations are relevant in Eq.~\eqref{eq:Pbeam}, and the jet target polarization can be well approximated by
\begin{equation}
  P_\text{jet}= \frac{|P_+| + |P_-|}{2} = \cos^2{\theta}-2\epsilon_\text{I}-\langle\epsilon\rangle,
  \label{eq:Pjet}
\end{equation}
where
\begin{equation}
    \langle\epsilon\rangle = \frac{\epsilon_{13}+\epsilon_{24}}{2}.
\end{equation}
Second-order corrections are omitted in Eq.~\eqref{eq:Pjet}.
Considering Eqs. \eqref{eq:P+} and \eqref{eq:P-} one can easily estimate that if
\begin{equation}
  |\nu|<2.5\%,\quad \epsilon_{13},\epsilon_{24},\epsilon_\text{I}<1\%,
  \label{eq:cond}
\end{equation}
the systematic uncertainty in the value of $P_\text{jet}$ due to omitting the second-order corrections is less than 0.1\%.

\subsection{The Breit--Rabi Polarimeter}

After passing through the separating and focusing magnet system~II, the hydrogen atoms are detected by an ion gauge (the BRP detector).

Depending on the jet polarization state ($P_+$, $P_-$, or $P_0$), the signals measured in the ion gauge are proportional to
\begin{align}
  m_+ &= (1-\nu') + (1+\nu')\epsilon_{24} + \epsilon_\text{I}' + \epsilon_\text{II}
  \nonumber \\
  &\approx 1 - \nu' + \epsilon_{24} + \epsilon_\text{I}' + \epsilon_\text{II}
  \approx 1,
  \label{eq:m+}  
\end{align}
\begin{align}
  m_- &= (1-\nu')\epsilon_{13} + (1+\nu') + \epsilon_\text{I}' + \epsilon_\text{II}
  \nonumber \\
  &\approx 1 + \nu' + \epsilon_{13} + \epsilon_\text{I}' + \epsilon_\text{II}
  \approx 1,
  \label{eq:m-}
\end{align}
\begin{align}
  m_0 &= (1-\nu')\epsilon_{13} + (1+\nu')\epsilon_{24}
  + 2\epsilon_\text{I}' + 2\epsilon_\text{II}
  \nonumber \\
  &\approx 2\left[\langle\epsilon\rangle + \epsilon_\text{I}' + \epsilon_\text{II}\right].
  \label{eq:m0}
\end{align}

Here, $\nu'$ and $\epsilon_\text{I}'$ are modified values of $\nu$ and $\epsilon_\text{I}$ that account for transmission efficiencies in the separating magnet system~II, and $\epsilon_\text{II}$ represents the inefficiency of suppressing the electron spin $-1/2$ states in this system. The second-order terms in Eqs.\,\eqref{eq:m+}--\eqref{eq:m0} can be safely neglected provided that $\nu'$, $\epsilon_\text{I}'$, and $\epsilon_\text{II}$ satisfy the conditions listed in Eq.\,\eqref{eq:cond}.

  For numerical estimates, the following typical BRP signal counts in the RHIC 2025 polarized proton run were used
  \begin{equation}
    m_+=1755,\quad m_-=1730,\quad m_0=10,\quad\tilde{m}_0=3500,
    \label{eq:counts}
  \end{equation}
  were $\tilde{m}_0$ is the BRP signal if both RF transitions are off.
Accordingly, 
\begin{equation}
  \frac{m_0}{m_+ + m_-}\approx%
  \langle\epsilon\rangle + \epsilon_\text{I}' + \epsilon_\text{II}%
  \approx 0.3\%
  \label{eq:m0/m+-}
\end{equation}
provides the upper limit for average inefficiency $\langle\epsilon\rangle$ of the RF transitions (as well as the upper limit for $\epsilon_\text{I}' + \epsilon_\text{II}$). 
Therefore, the polarization determined as
\begin{equation}
  P_\text{BRP} = \cos^2\theta - \frac{m_0}{m_+ + m_-},
  \label{eq:P_BRP}
\end{equation}
yields an unbiased and precisely determined estimate of the jet target polarization,
\begin{equation}
  P_\text{BRP} - P_\text{jet} = 2\epsilon_\text{I} - \epsilon_\text{I}' - \epsilon_\text{II},
\end{equation}
under the simplifying assumption that 
$\epsilon_\text{I} = \epsilon_\text{I}' = \epsilon_\text{II}$
holds or, in particular, that both separating magnet systems operate with
100\% efficiency.

It is important to note that the separation-magnet inefficiencies
$\epsilon_\text{I}$, $\epsilon_\text{I}'$, and $\epsilon_\text{II}$
cannot be directly monitored by either the recoil spectrometer or the BRP.
Consequently, establishing a reliable relation between
$P_\text{BRP}$ and $P_\text{jet}$
requires detailed simulations of hydrogen atom trajectories through the
complete HJET magnetic system.

\subsection{Discussion}

In evaluating $P_\text{BRP}$, we did not consider measurements with both RF transitions
switched  off. In this case, the BRP signal may be approximated as
\begin{equation}
  \widetilde{m}_0 = (1-\nu') + (1+\nu') = 2.
  \label{eq:~m0}
\end{equation}
Since
\begin{equation}
  m_+ + m_- = m_0 + \widetilde{m}_0,
  \label{eq:+-=00}
\end{equation}
the measurement of $\widetilde{m}_0$ cannot improve the accuracy of the BRP determination of $P_\text{jet}$. Using \eqref{eq:counts}, one finds
\begin{equation}
  \langle\epsilon\rangle + \epsilon_\text{I}' + \epsilon_\text{II} =
  \frac{m_+ + m_- - \widetilde{m}_0}{m_+ + m_-}
  \approx -0.43\%,
\end{equation}
whereas
\begin{equation}
  \langle\epsilon\rangle + \epsilon_\text{I}' + \epsilon_\text{II} =
  \frac{m_0}{m_+ + m_-}
  \approx 0.29\%.
  \label{eq:m0_meas}
\end{equation}
The discrepancy, which is significant compared to the value of $m_0$, may be explained by a small nonlinearity, at the level of approximately $0.7\%$, in the response of the ion gauge to the measured atomic hydrogen flux.

The original RHIC HJET design~\cite{Wise:2004uy} assumed the use of a second RF unit (located in chamber~8) for BRP calibration. However, it was later found that, owing to the small observed value of $\langle\epsilon\rangle$, this additional RF unit is not required~\cite{ZelenskiPrivate2022} to determine the jet target polarization with an accuracy of approximately $0.1\%$. Assuming $\epsilon_\text{I} = \epsilon_\text{I}' = \epsilon_\text{II} = 0$ and using Eq.\,\eqref{eq:m0_meas}, together with another recent BRP measurement,
\begin{equation}
  \nu' + \frac{\epsilon_{13} - \epsilon_{24}}{2}
  = \frac{m_- - m_+}{m_+ + m_-}
  \approx 0.7\%,
\end{equation}
one readily finds that the values of $\nu'$, $\epsilon_{13}$, and $\epsilon_{24}$ satisfy the conditions~\eqref{eq:cond}. Consequently, the equality $P_\text{BRP} = P_\text{jet}$ is achieved with sufficient accuracy.

This conclusion remains valid in the more general case in which
\begin{equation}
  2\epsilon_\text{I} \approx \epsilon_\text{I}' + \epsilon_\text{II}.
\end{equation}
However, since the BRP measurements are not sensitive to the value of $\epsilon_\text{I}$, one must rely on simulations of atomic hydrogen tracking in the HJET system. In this context, BRP measurements performed with both SFT and WFT transitions turned on in both RF units,
\begin{equation}
  m_0' = 2\epsilon_\text{II},
\end{equation}
may be particularly useful for tuning and validating the simulation software.

In Ref.\,\cite{Wise:2006xj}, it was suggested to use atomic beam blockers to completely reject states $|3\rangle$ and $|4\rangle$. For the estimates presented here, such rejection corresponds to $\epsilon_\text{I}' \to 0$ and $\epsilon_\text{II} \to 0$. As a result, the BRP would measure the unbiased RF-transition inefficiency $\langle\epsilon\rangle$. However, it was not fully appreciated that the effective jet polarization in the scattering chamber remains affected by $\epsilon_\text{I}$, leading to a biased BRP evaluation of the jet target polarization when $\epsilon_\text{I} \neq 0$.

Therefore, the installation of atomic beam blockers does not necessarily improve the accuracy of the determination of $P_\text{jet}$. Once again, this emphasizes the importance of accurate simulations of atomic hydrogen tracking. From this perspective, a comparison of BRP measurements performed with and without blockers may be of critical importance.

Admixture of molecular hydrogen in the polarized atomic jet effectively dilutes the jet target polarization; however, this component may be invisible to the BRP. Assuming that the dominant source of H$_2$ in the scattering chamber is recombination of defocused jet atoms in chambers 5 and 7, followed by diffusion into chamber 6, one may expect an approximately uniform density of molecular hydrogen in the scattering region. Since the $z$-coordinate profile of the target density is routinely monitored in the HJET data analysis, the H$_2$ component can be subtracted from the atomic jet data in a straightforward manner~\cite{Poblaguev:2020qbw}.

However, this method is not applicable to H$_2$ originating from the nozzle. In this case, the corresponding dilution of the jet target polarization was evaluated in a dedicated measurement with the dissociator turned off~\cite{Poblaguev:2020qbw}. The effect was found to be negligible, $\lesssim0.1\%$. Accordingly, the possible contribution of this molecular hydrogen to the BRP-measured value given in Eq.\,\eqref{eq:m0/m+-} is expected to be small. Nevertheless, it would be worthwhile to evaluate this nozzle-originating H$_2$ contribution using a jet-tracking simulation.

\section{Beam-induced depolarization}
\label{sec:Meth}

In the analysis of beam-induced depolarization, we use the EIC flattop beam parameters listed in Table~\ref{tab:BeamPar}.

\subsection{Bunch-induced magnetic field}

Assuming a Gaussian longitudinal bunch density, the temporal structure of the EIC polarized proton beam current, $I_b(t)$, can be approximated as
\begin{equation}
  I_b(t) = I_\text{pk}
  \sum_{k=-\infty}^{\infty}
  \exp\!\left[-\frac{(t-k\tau_b)^2}{2\sigma_t^2}\right],
  \label{eq:Ibeam}
\end{equation}
where $\tau_b$ is the bunch spacing, $\sigma_t$ is the rms bunch length in time, and
  $I_\text{pk} = e N_p/(\sqrt{2\pi}\,\sigma_t)\approx22\,\text{A}$ \cite{Rathmann:2025jgp} is the peak current.

For round beam, the induced magnetic field can be readily calculated using the Amp\`ere--Maxwell law, stating that the circulation of the magnetic field $\boldsymbol{B}$ is determined by the total enclosed current $I$,
\begin{equation}
  \oint \boldsymbol{B}\cdot d\boldsymbol{s} = \mu_0 I,
  \label{eq:AM}
\end{equation}
where $\mu_0 = 4\pi\times10^{-7}\,\text{T\,m/A}$ is the vacuum magnetic permeability.

\begin{figure}[t]
\begin{center}
\includegraphics[width=0.95\columnwidth]{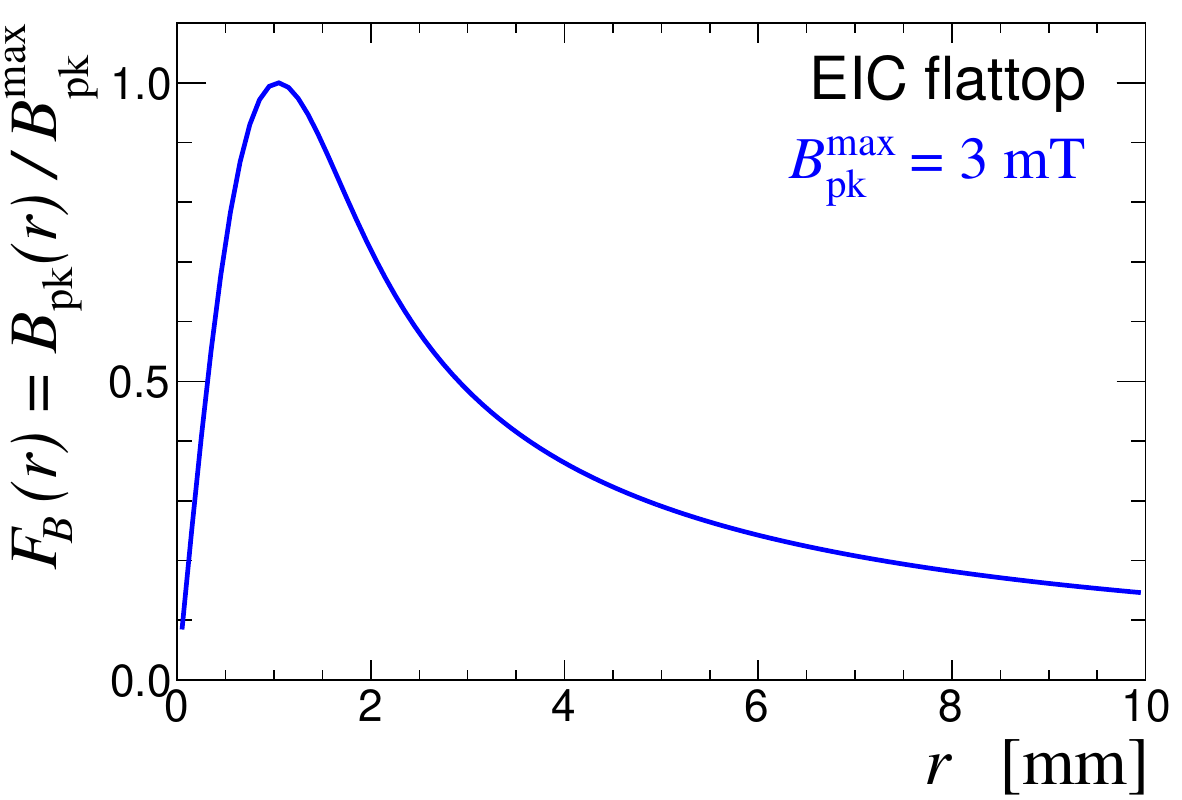}%
\end{center}
\caption{\label{fig:BunchField}
  Unity-normalized radial profile of the beam-induced magnetic field at the EIC flattop.}
\end{figure}

For a Gaussian transverse density of the beam current with rms $\sigma_r$, the current enclosed in a circle of radius $r$ can be easily calculated and one obtains for the peak value of the beam induced magnetic field $B_\text{ind}(r,t)$ 
\begin{equation}
  B_\text{pk}(r) = B_\text{ind}(r,0) 
  = \frac{\mu_0 I_\text{pk}}{2\pi r}
    \left(1 - e^{-r^2/2\sigma_r^2}\right)
    = B_\text{pk}^\text{max} F_B(r),
    \label{eq:Bpk}
\end{equation}
where $B_\text{pk}^\text{max} \approx 3\,\text{mT}$ for the EIC flattop beam, and
$F_B(r)$ is a unity-normalized radial profile shown in Fig.~\ref{fig:BunchField}.

\begin{figure}[t]
\begin{center}
\includegraphics[width=0.95\columnwidth]{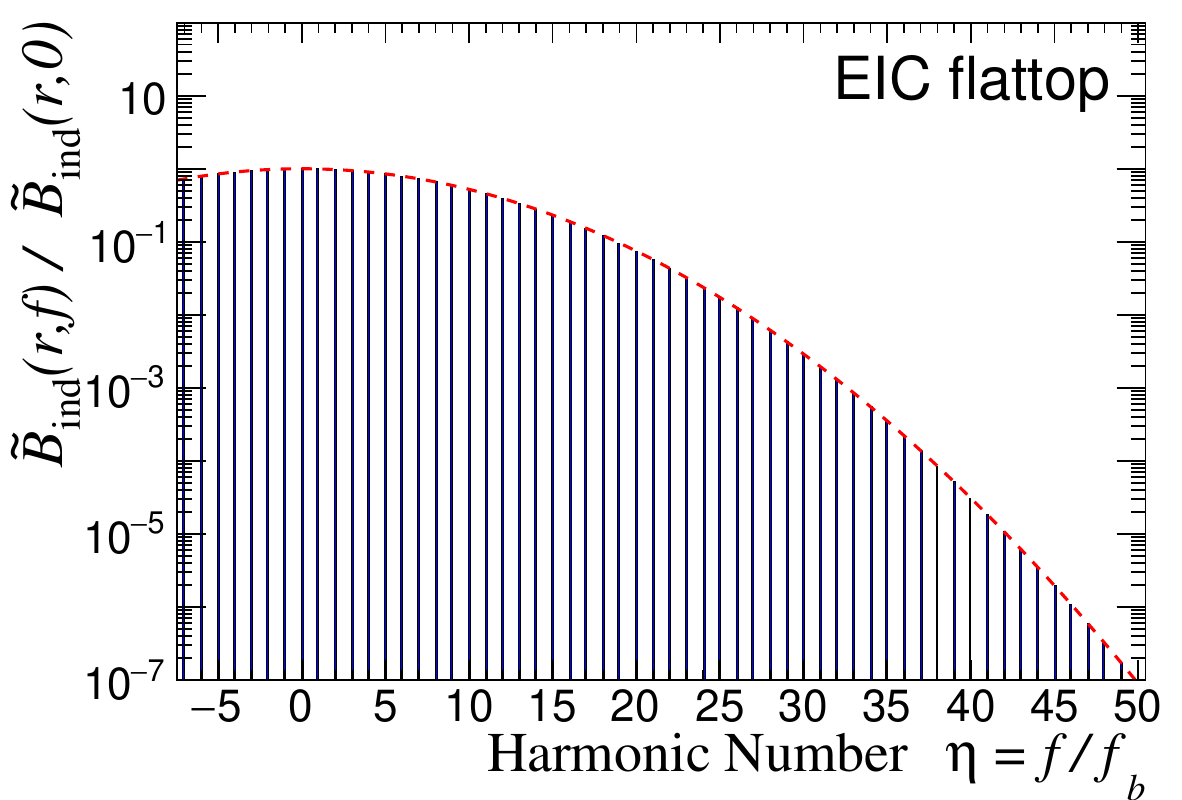}
\end{center}
\caption{\label{fig:BunchHarm}
  Fourier spectrum of the proton beam longitudinal profile planned for the EIC.
  The red dashed line represents the Gaussian envelope defined in Eq.\,\eqref{eq:sf}.}
\end{figure}

The Fourier transform $B_\text{ind}(r,t) \to \tilde{B}_\text{ind}(r,f)$ yields a discrete, equidistant set of $\delta$-function harmonics with spacing $f_b = 1/\tau_b$, modulated by a Gaussian spectral envelope, as illustrated in Fig.~\ref{fig:BunchHarm}. The rms width of this envelope is
\begin{equation}
  \sigma_f = \frac{1}{2\pi\sigma_t}
  \approx 796\,\text{MHz}.
  \label{eq:sf}
\end{equation}

Consequently, the beam-induced magnetic field at radius $r$ can be expressed as a sum of harmonic components,
\begin{align}
  B_\text{ind}(r,t) 
  &= B_\text{osc}^\text{max} F_B(r) 
  \sum_{\eta=-\infty}^{\infty} e^{-(\omega_\eta \sigma_t)^2/2}e^{i\omega_\eta t}
  \nonumber \\
  &= B_\text{osc}^\text{max} F_B(r)
  \left[
    1 + 2\sum_{\eta=1}^{\infty}e^{-(\omega_\eta \sigma_t)^2/2}
    \cos(\omega_\eta t)
  \right],
  \label{eq:Bind}
\end{align}
where $\omega_\eta = 2\pi \eta f_b$, and
\begin{equation}
  B_\text{osc}^\text{max}
  = B_\text{pk}^\text{max}\,\frac{\sqrt{2\pi}\sigma_t}{\tau_b}
  \approx 0.136\,\text{mT}.
  \label{eq:zero}
\end{equation}
The normalization factor $\sqrt{2\pi}\sigma_t/\tau_b$ can be readily determined by noting that the zeroth harmonic amplitude is fixed by the average beam current.

\subsection{Matrix elements for bunch-induced transitions}

For a hydrogen atom subjected to a weak oscillating magnetic field
$\boldsymbol{B}_\omega \cos(\omega t)$,
the perturbation enters the Hamiltonian as
\begin{equation}
  H_1(t) = -\mu_e \left[
    B_\parallel
    \begin{pmatrix}
      1 & 0 \\
      0 & -1
    \end{pmatrix}_e
    +
    B_\perp
    \begin{pmatrix}
      0 & 1 \\
      1 & 0
    \end{pmatrix}_e
  \right]
  \cos(\omega t),
\end{equation}
where $B_\parallel$ and $B_\perp$ are the components of the oscillating magnetic field
parallel and perpendicular to the holding field, respectively.
The Pauli matrices act only on the electron spin.

The parallel component $B_\parallel$ induces $\sigma$ transitions
($\Delta F=\pm1$, $\Delta m_F=0$).
Following Ref.~\cite{Beijers:2005}, the corresponding matrix elements can be written as
\begin{equation}
  {\cal M}_{ij}^\sigma = \mu_{ij}^\sigma B_\parallel, \quad
  \mu_{ij}^\sigma = 2\mu_e
  \begin{pmatrix}
    1 & 0              & 0  & 0              \\
    0 & \cos 2\theta   & 0  & -\sin 2\theta  \\
    0 & 0              & -1 & 0              \\
    0 & -\sin 2\theta  & 0  & -\cos 2\theta
  \end{pmatrix}.
  \label{eq:Msigma}
\end{equation}

The appearance of diagonal matrix elements in Eq.~\eqref{eq:Msigma}
can be understood by considering the zeroth harmonic of the beam-induced field,
corresponding to the average longitudinal field
$\langle B_\text{ind}^\parallel\rangle$.
For example, retaining only the term
${\cal M}_{11}=\mu_{11}^\sigma\langle B_\text{ind}^\parallel\rangle$,
Eq.~\eqref{eq:da/dt} reduces to
\begin{equation}
  i\hbar\,d\ln a_1 = \int {\cal M}_{11}(t)\,dt.
\end{equation}
This results in the following modification of the phase of the $|1\rangle$ state:
\begin{equation}
  \exp\!\left(\frac{iE_{|1\rangle} t}{\hbar}\right)
  \;\longrightarrow\;
  \exp\!\left(
    \frac{iE_{|1\rangle} t}{\hbar}
    - \frac{i2\mu_e\langle B_\text{ind}^\parallel\rangle t}{\hbar}
  \right),
\end{equation}
which reflects a shift of the $|1\rangle$ energy level due to an effective change
in the holding field caused by the zeroth-harmonic component of the beam-induced field.

\begin{figure}[t]
\begin{center}
\includegraphics[width=0.7\columnwidth]{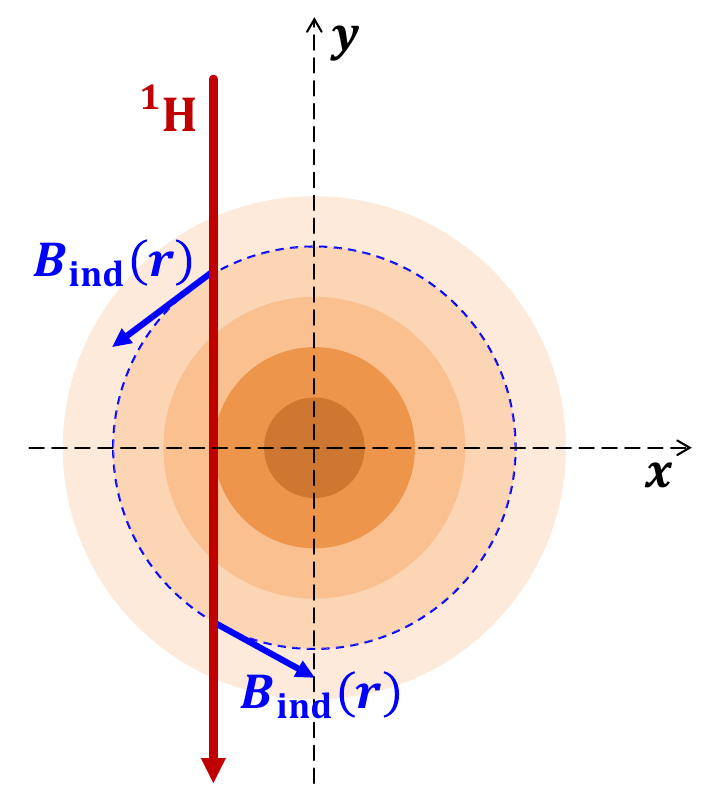}
\end{center}
\caption{\label{fig:Track}
  Tracking of a hydrogen atom through the $z$-directed proton beam.}
\end{figure}

Therefore, the diagonal matrix elements in Eq.~\eqref{eq:Msigma}
can be neglected, provided that the time-dependent variation of the effective holding field,
\begin{equation}
  B_\text{hold}^\text{eff}(r,t)
  = B_\text{hold}
    + B_\text{ind}(r,t)\frac{x}{r},
\end{equation}
(where the coordinate $x$ is defined in Fig.~\ref{fig:Track}),
is explicitly taken into account in the instantaneous eigenenergies
$E_{|i\rangle}(t)$ and in the mixing angle $\theta(t)$.
This procedure corresponds to using time-dependent Rabi frequencies
$\omega_R(t)$ and detunings
$\Delta\omega(t)=\omega-\omega_{fi}(t)$
in Eq.~\eqref{eq:alpha}.

Notably, the variation of the atomic energy levels is driven mainly by a coherent sum
of a large number of low-frequency harmonics, whereas resonant depolarization
(for the HJET holding field of 120\,mT) can be induced only by a single
high-frequency harmonic. As a result, these two processes can be treated independently.

For $\pi$ transitions
($\Delta F=0,1$ and $\Delta m_F=\pm1$),
induced by the transverse oscillating field $B_\perp$,
the matrix elements are
\begin{equation}
  {\cal M}_{ij}^\pi = \mu_{ij}^\pi B_\perp, \quad
  \mu_{ij}^\pi = 2\mu_e
  \begin{pmatrix}
    0             & \sin\theta & 0              & \cos\theta \\
    \sin\theta    & 0          & \cos\theta     & 0          \\
    0             & \cos\theta & 0              & -\sin\theta \\
    \cos\theta    & 0          & -\sin\theta    & 0
  \end{pmatrix}.
  \label{eq:Mpi}
\end{equation}

\subsection{Tracking of hydrogen atoms through the beam region}

For an accurate evaluation of beam-induced transition probabilities,
the trajectory of a hydrogen atom through the beam-induced magnetic field
must be taken into account.
If, at $t=0$, the atomic coordinates are $x_\text{at}(0)$ and $y_\text{at}(0)$
(see Fig.~\ref{fig:Track}),
and the atom moves vertically with velocity $v_\text{at}$,
its coordinates and radial distance from the beam axis at time $t$ are given by
\begin{align}
  x_\text{at}(t) &= x_\text{at}(0), \\
  y_\text{at}(t) &= y_\text{at}(0) - v_\text{at} t, \\
  r_\text{at}(t) &= \sqrt{x_\text{at}^2(t) + y_\text{at}^2(t)} .
\end{align}

Considering a transition $|i\rangle\!\to\!|f\rangle$ driven by the $\eta$th beam harmonic
of the beam-induced magnetic field
($\omega_\eta = 2\pi \eta f_b$),
the instantaneous driving frequencies along the atomic trajectory are
\begin{align}
  \omega_R(t) &=
  \frac{B_\text{osc}^\text{max} F_B\!\left(r_\text{at}(t)\right)}
       {\hbar\, r_\text{at}(t)}
  e^{-2(\pi\eta\sigma_t/\tau_b)^2}
  \begin{cases}
    \mu_{fi}^\sigma\, x_\text{at}(t), & \text{$\sigma$ transitions}, \\
    \mu_{fi}^\pi\, y_\text{at}(t),    & \text{$\pi$ transitions},
  \end{cases}
  \label{eq:omegaR}
  \\
  \Delta\omega(t) &=
  \omega_\eta
  - 2\pi f_{fi}\!\left(
      B_\text{hold}^\text{eff}\!\left(r_\text{at}(t),t\right)
    \right),
  \label{eq:omegaD}
\end{align}
where $f_{fi}(B_\text{hold}^\text{eff})$ is the transition frequency
determined by the instantaneous effective holding field.

Substituting $\omega_R(t)$ and $\Delta\omega(t)$ into
Eqs.~\eqref{eq:alpha},
one can calculate the evolution of the probability amplitudes
$\alpha_i(t)$ and $\alpha_f(t)$
along the atomic trajectory through the beam.

In the HJET geometry, the region of significant beam-induced magnetic field
is limited by the scattering chamber height,
$L_\text{int} = 60\,\text{mm}$.
Assuming that the $z$-directed proton beam axis is located at
$x_\text{beam}=y_\text{beam}=0$,
the initial vertical coordinate of the hydrogen atom is taken as
\begin{equation}
  y_\text{at}(0) = +\frac{L_\text{int}}{2},
\end{equation}
and the tracking is terminated when the atom reaches
\begin{equation}
  y_\text{at}(t) = -\frac{L_\text{int}}{2}.
\end{equation}

The horizontal coordinate $x_\text{at}(0)$ of the atoms is distributed
according to the jet density profile~\cite{Poblaguev:2020qbw},
\begin{equation}
  \frac{dN}{dx} \propto
  \exp\!\left(-\frac{x^2}{2\sigma_\text{jet}^2}\right),
  \label{eq:x}
\end{equation}
with $\sigma_\text{jet}\approx 2.6\,\text{mm}$.

According to the evaluation of the atomic velocity distribution
in Ref.~\cite{Wise:2006xj},
the vertical velocity $v_\text{at}$ follows
\begin{equation}
  \frac{dN}{dv_\text{at}} \propto
  v_\text{at}^2
  \exp\!\left[
    -\frac{\left(v_\text{at}-v_\text{drift}\right)^2}{2\sigma_v^2}
  \right],
  \label{eq:v}
\end{equation}
where $v_\text{drift}\approx 1800\,\text{m/s}$
and $\sigma_v\approx 400\,\text{m/s}$.

A typical interaction time can thus be estimated as
\begin{equation}
  t_\text{int} =
  \frac{L_\text{int}}{v_\text{drift}}
  \approx 33\,\mu\text{s}.
\end{equation}

\section{Estimated depolarization of the HJET target at EIC}
\label{sec:Est}

\begin{table}[t]
  \caption{\label{tab:EIC-129}
    Transition frequencies (in units of $f_b$) for the EIC flattop beam
    as functions of the holding field $B_\text{hold}$.
    The last column shows the relative suppression factor due to the Gaussian
    spectral envelope, evaluated at $B_\text{hold}=129\,\text{mT}$.
  }
  \begin{center}
    \begin{tabular}{l | r | r | r | r}
      & \multicolumn{3}{c|}{$f_{ij}/f_b$}
      & \multicolumn{1}{c}{$\exp(-f_{ij}^2/2\sigma_f^2)$} \\[5pt]
      & \multicolumn{1}{c|}{126\,mT}
      & \multicolumn{1}{c|}{129\,mT}
      & \multicolumn{1}{c|}{132\,mT}
      & \multicolumn{1}{c}{129\,mT} \\[3pt]
      \hline & & & & \\[-7pt]
      $f_{12}^\pi$       &  6.316 &  6.348 &  6.380 & $7.70\times10^{-1}$ \\
      $f_{34}^\pi$       &  9.348 &  9.315 &  9.284 & $5.69\times10^{-1}$ \\
      $f_{23}^\pi$       & 32.624 & 33.518 & 34.414 & $6.79\times10^{-4}$ \\
      $f_{13}^{2\gamma}$ & 38.940 & 39.867 & 40.794 & $3.30\times10^{-5}$ \\
      $f_{24}^\sigma$    & 41.972 & 42.834 & 43.698 & $6.71\times10^{-6}$ \\
      $f_{14}^\pi$       & 48.288 & 49.182 & 50.078 & $1.51\times10^{-7}$
    \end{tabular}
  \end{center}
\end{table}

Resonant depolarization may occur when a harmonic $\eta$ of the beam-induced
magnetic field coincides with a hyperfine transition frequency,
\begin{equation}
  f_{ij}(B_\text{hold}) = \eta f_b,
\end{equation}
which imposes a constraint on the choice of the holding field
$B_\text{hold}$.

Using Eqs.~\eqref{eq:f}, one readily finds that, for the EIC parameters,
a change of the holding field by only
$\Delta B_\text{hold}=3.2\,\text{mT}$
shifts the transition frequency $f_{13}$ by one harmonic unit, $f_b$.
A similar sensitivity to the holding field is observed for the
$f_{14}$, $f_{23}$, and $f_{24}$ transitions.

Since the beam-induced magnetic field modifies the effective holding field
by up to $\pm3\,\text{mT}$ (depending on the transverse coordinates $x$ and $y$),
some hydrogen atoms will inevitably satisfy resonant transition conditions
while traversing the scattering chamber, as illustrated in
Table~\ref{tab:EIC-129}.
The holding field value $B_\text{hold}=129\,\text{mT}$ considered for the EIC
differs from the $120\,\text{mT}$ used at RHIC.
As demonstrated below, $B_\text{hold}=129\,\text{mT}$ corresponds to a local
minimum of the overall beam-induced depolarization.

Nevertheless, even at zero detuning, the depolarization transition probability
given by Eq.~\eqref{eq:Prob} may remain small.

This behavior is due to the suppression of the Rabi frequency $\omega_R$
[see Eq.~\eqref{eq:omegaR}] by the exponential factor
$\exp(-f_{ij}^2/2\sigma_f^2)$,
as well as to the possible smallness of the effective resonance time
$t_\text{res}\ll t_\text{int}$.
The latter arises from the explicit dependence of the beam-induced magnetic
field $B_\text{ind}(r,t)$ on both the atomic coordinates and time.
Consequently, a reliable evaluation of the
$|i\rangle\!\to\!|f\rangle$ transition probability
(assuming $\alpha_i(0)=1$ and $\alpha_f(0)=0$)
requires an explicit time-dependent calculation of $a_f(t)$ along the
hydrogen atom trajectory.

In such a study, three transition probabilities are of particular interest:

\noindent\textbf{$w_{ij}^\text{brp}$} --- the transition probability
$|a_f(t)|^2$ evaluated at the exit of the interaction region,
$y=-L_\text{int}/2$.
This quantity is relevant for estimating the depolarization correction to the
polarization determined by the Breit--Rabi polarimeter (BRP).

\noindent\textbf{$w_{ij}^\text{jet}$} --- the transition probability
$|a_f(t)|^2$ convoluted with the transverse proton beam density profile.
It provides an estimate of the effective depolarization of the jet target
relevant for the proton beam polarization measurement.

\noindent\textbf{$w_{ij}^\text{max}$} --- the maximum value of $|a_f(t)|^2$
attained along the hydrogen atom trajectory through the region of the
beam-induced magnetic field.
The quantity $(w_{ij}^\text{max})^{1/2}$ may be used as a practical criterion
for determining whether the transition
$|i\rangle\!\to\!|j\rangle$ can be safely neglected in the system of
equations~\eqref{eq:da/dt}.

\begin{table*}[t]
  \caption{\label{tab:23}
    Dependence of the average $f_{23}^\pi$ transition probabilities
    $w_{23}^\text{brp}$, $w_{23}^\text{jet}$, and $w_{23}^\text{max}$
    on the holding field $B_\text{hold}$, the initial transverse coordinate
    $x(0)$ of the hydrogen atom, and its velocity $v_\text{at}$.
    The value $B_\text{hold}=130.6137\,\text{mT}$ corresponds to the exact
    resonant condition $f_{23}=34f_b$ for the $|2\rangle\!\to\!|3\rangle$
    transition in the absence of beam-induced fields.
    A reference to an equation in the table indicates that the corresponding
    parameter is distributed according to that equation.
  }
  \begin{center}
    \begin{tabular}{r | c c c | c c c}
      & $B_\text{hold}$ [mT] & $x(0)$ & $v_\text{at}$ &
      $w_{23}^\text{brp}$ & $w_{23}^\text{jet}$ & $w_{23}^\text{max}$ \\[3pt]
      \hline \\[-7pt]
      1 & 130.6137 & 0                 & $v_\text{drift}$   &
          $1.4\times10^{-9}$ & $3.6\times10^{-4}$ & $4.0\times10^{-4}$ \\
      2 & 130.6137 & 0                 & Eq.~\eqref{eq:v}   &
          $1.6\times10^{-9}$ & $3.4\times10^{-4}$ & $3.8\times10^{-4}$ \\
      3 & 130.6137 & Eq.~\eqref{eq:x}  & $v_\text{drift}$   &
          $1.2\times10^{-4}$ & $1.0\times10^{-4}$ & $1.4\times10^{-4}$ \\
      4 & 129      & Eq.~\eqref{eq:x}  & $v_\text{drift}$   &
          $5.8\times10^{-12}$ & $2.9\times10^{-10}$ & $5.9\times10^{-10}$
    \end{tabular}
  \end{center}
\end{table*}

Although $w_{ij}^\text{brp}$, $w_{ij}^\text{jet}$, and $w_{ij}^\text{max}$
are referred to below as depolarization probabilities,
the actual change in the population (and polarization) of the atomic state
$|i\rangle$ is given by
\begin{equation}
  P_{|i\rangle}
  \;\to\;
  P_{|i\rangle}(1-w_{ij}) + P_{|j\rangle}w_{ij}
  \;=\;
  P_{|i\rangle}
  + w_{ij}\left(P_{|j\rangle}-P_{|i\rangle}\right).
\end{equation}

Notably, $w_{24}=0.5$ corresponds to complete depolarization of a proton for
an atom initially in the state $|2\rangle$,
whereas $w_{24}=1$ corresponds to a full proton spin flip.
In contrast, for the transition $2\to3$, the case $w_{23}=1$ changes the proton
polarization from $-\cos 2\theta$ to $-1$.

\subsection{Potentially resonant transitions  \label{sec:PotRes}}

According to Table~\ref{tab:EIC-129}, three transitions,
$f_{23}^\pi$, $f_{24}^\sigma$, and $f_{14}^\pi$, can be regarded as
\emph{potentially resonant}, meaning that an exact resonance condition
may be satisfied while a hydrogen atom traverses the proton beam.
The $f_{13}$ transition is not considered here, since it corresponds to
$\Delta m_F = 2$ and is forbidden within the dipole approximation adopted
in this work.

Assuming an exactly resonant holding field $B_\text{hold}^\text{res}$,
such that $f_{ij}(B_\text{hold}^\text{res}) = \eta f_b$ (prior to beam-induced corrections),
and using the maximum possible Rabi frequency,
\begin{equation}
  \omega_R^\text{max} =
  \frac{\mu_{ij} B_\text{osc}^\text{max}}{\hbar}
  \exp\!\left[-2(\pi\eta\sigma_t/\tau_b)^2\right],
  \label{eq:omegaR_max}
\end{equation}
one obtains an upper bound on the depolarization transition probability,
\begin{equation}
  w_{ij}^\text{upper}
  = \left(\frac{\omega_R^\text{max} t_\text{int}}{2}\right)^2.
  \label{eq:upper}
\end{equation}

For the $f_{23}^\pi$ transition,
$B_\text{hold}^\text{res} = 130.6137\,\text{mT}$, yielding
\begin{equation}
  w_{23}^\text{upper} \simeq 5\times10^{-2},
\end{equation}
which, by itself, does not exclude a substantial resonant transition probability.

A more accurate analysis, including the dependence on the initial atomic
coordinate $x(0)$, atomic velocity $v_\text{at}$, and the nominal holding
field $B_\text{hold}$, as well as on
the variations of $\omega_R$ and $\Delta\omega$ along the atomic trajectory,
is summarized in Table~\ref{tab:23}.

For the resonant holding field and trajectories with $x(0)=0$
(lines~1 and~2 of Table~\ref{tab:23}),
the values of $w_{23}^\text{brp}$ are extremely small and are consistent
with the numerical accuracy of the calculation.
Along such trajectories, the beam-induced magnetic field is always
horizontal and does not modify the $f_{23}^\pi$ transition frequency,
maintaining $\Delta\omega=0$.
Moreover, due to the antisymmetry of the induced field with respect to the
$y=0$ axis, the net effect of the evolution governed by
Eqs.~\eqref{eq:alpha} vanishes identically for atomic motion from
$y=L_\text{int}/2$ to $y=-L_\text{int}/2$.

Lines~1 and~2 also demonstrate that incorporating the atomic velocity
distribution does not significantly alter the estimated depolarization
relative to calculations performed with a fixed velocity
$v_\text{at}=v_\text{drift}$.
Consequently, all subsequent simulations were performed using a fixed
atomic velocity in order to reduce computational time.

\begin{figure}[t]
\begin{center}
\includegraphics[width=0.95\columnwidth]{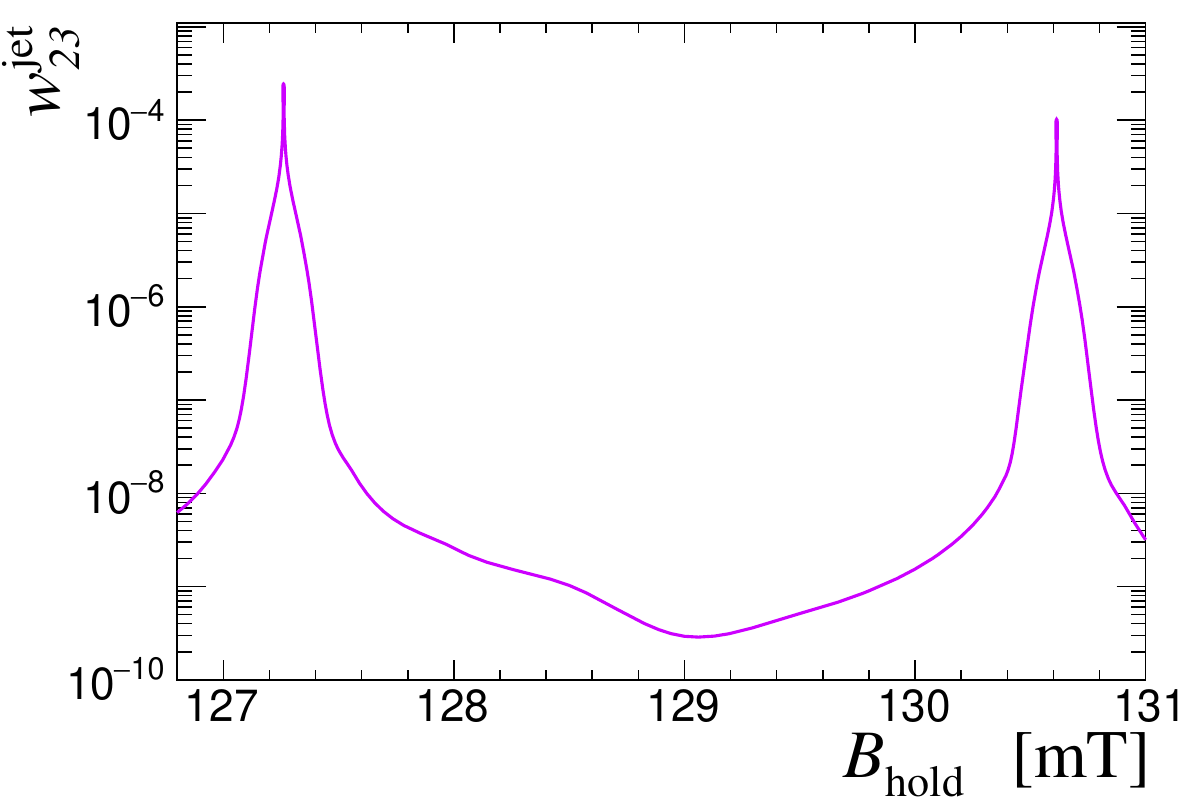}
\end{center}
\caption{\label{fig:23}
  The recoil-spectrometer-observed $|2\rangle \to |3\rangle$ transition
  probability as a function of the holding magnetic field.}
\end{figure}

Including the jet density profile results in transition probabilities of
order ${\cal O}(10^{-4})$ for $w_{23}^\text{brp}$,
$w_{23}^\text{jet}$, and $w_{23}^\text{max}$, even for the resonant holding
field.
A modest deviation of $B_\text{hold}$ from the resonant value suppresses
the transition probability to a completely negligible level, as illustrated
in Fig.~\ref{fig:23}.

For the remaining \emph{potentially resonant} transitions,
$f_{24}^\sigma$ and $f_{14}^\pi$, negligible beam-induced depolarization
follows directly from the upper-limit estimate~\eqref{eq:upper}:
\begin{align}
  w_{24}^\text{upper} &= 8\times10^{-7},
  \quad (B_\text{hold}=129.5785\,\text{mT},~\eta=43),
  \label{eq:upper24} \\
  w_{14}^\text{upper} &= 5\times10^{-9},
  \quad (B_\text{hold}=128.3898\,\text{mT},~\eta=49).
\end{align}

Thus, none of the \emph{potentially resonant} transitions can produce a measurable
change in the jet target polarization at the EIC, and all such transitions
can be safely excluded from the system of equations~\eqref{eq:da/dt}.

\subsection{Non-resonant transitions}

With only two remaining (\emph{non-resonant}) transitions,
$f_{12}^\pi$ and $f_{34}^\pi$, the four-level system
\eqref{eq:da/dt} naturally decomposes into two independent two-level
subsystems. In this case, evaluation of $w_{ij}^\text{max}$ is no longer required.

Applying Eq.\,\eqref{eq:alpha-} to these \emph{non-resonant} transitions,
one finds that in both cases the generalized Rabi frequency $\Omega$ can be treated as
approximately constant, $\Omega/2 \approx 95\,\text{MHz}$, while the
instantaneous Rabi frequency satisfies
$|\omega_R(t)| \lesssim 3\,\text{MHz}$ and varies along the atomic trajectory.
Let $\tilde{\omega}_R(\Delta t)=\omega(t-t_0)$, where $t_0$ denotes the time at which the atomic
trajectory crosses the $y = 0$ plane. Then $\tilde{\omega}_R(\Delta t)$ is an odd
function of $\Delta t$, and for trajectories with $x(t) = 0$,
\begin{equation}
  |\tilde{\omega}_R(\Delta t)| \propto F_B(v_\text{at}|\Delta t|).
\end{equation}

Using Fig.\,\ref{fig:BunchField}, one can readily estimate the characteristic
Fourier frequencies $\Omega_R$ associated with the temporal variation of
$\omega_R(t)$:
\begin{equation}
  \Omega_R \lesssim \frac{\pi}{2}\,\frac{v_\text{at}}{\text{mm}}
  \approx 3\,\text{MHz}
  \ll \frac{\Omega}{2}.
\end{equation}
Thus, $\omega_R(t)$ may be regarded as a slowly varying function of time, which
justifies the use of Eq.\,\eqref{eq:alpha-} as a reasonable approximation for
the pre-evaluation of the \emph{non-resonant} transition probability
$w_\text{non-res}^\text{brp}$.
Within this approach,
\begin{equation}
  w_{\text{non-res}}^\text{brp}
  = \frac{\omega_R^2(L_\text{int}/v_\text{at})}{2\Omega^2}
  \approx 0.
\end{equation}

A more accurate estimate, based on numerical solutions of
Eqs.\,\eqref{eq:alpha} with time-dependent $\omega_R(t)$ and
$\delta\omega(t)$, yields
\begin{equation}
  w_\text{non-res}^\text{brp}
  \sim 0.02\, w_\text{non-res}^\text{jet}.
\end{equation}

Thus, acknowledging that the Breit--Rabi polarimeter does not possess sufficient
sensitivity to detect depolarization effects arising from the
\emph{non-resonant} transitions $f_{12}^\pi$ and $f_{34}^\pi$, we focus below on
the corresponding jet-target depolarization relevant for recoil spectrometer
measurements.

\begin{figure}[t]
\begin{center}
\includegraphics[width=0.95\columnwidth]{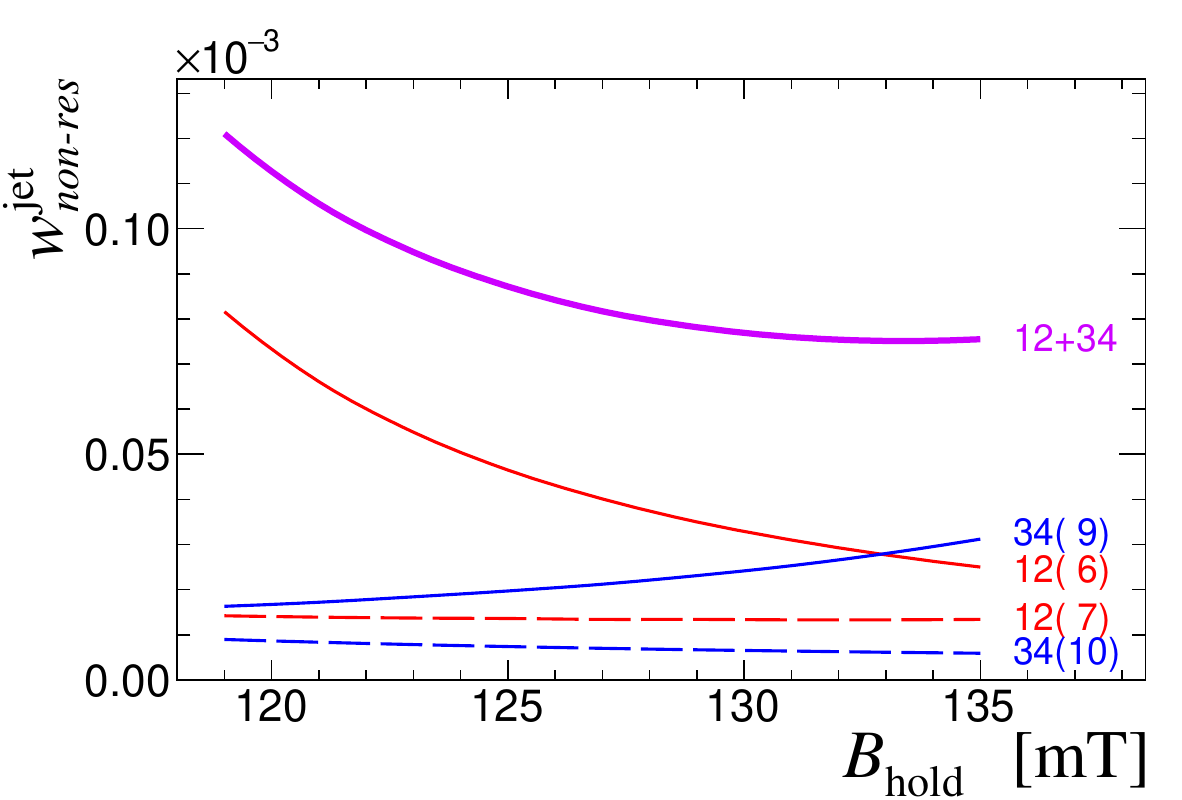}
\end{center}
\caption{Calculated transition probabilities for the \emph{non-resonant}
$f_{12}^\pi$ and $f_{34}^\pi$ transitions as functions of the holding magnetic
field. The transitions are labeled by the initial/final state $if$ and harmonic $(h)$ numbers. ``12+34'' means the sum of all four \emph{non-resonant} probabilities.}
\label{fig:12-34}
\end{figure}

For each transition, two neighboring harmonics $\eta$ and $\eta+1$ satisfying
\begin{equation}
  \eta < \frac{f_{ij}}{f_b} < \eta+1
\end{equation}
were taken into account. The results are presented in
Fig.\,\ref{fig:12-34}.

The total jet-target depolarization, as observed in recoil spectrometer
measurements,
\begin{equation}
  \Delta_\text{dep}|P_\pm|
  \approx -w_\text{non-res}^\text{jet}
  = -w_{12(6)}^\text{jet} -w_{12(7)}^\text{jet} -w_{34(9)}^\text{jet} -w_{34(10)}^\text{jet},
\end{equation}
was found to be approximately $-0.008\%$ for the HJET holding field
$B_\text{hold} = 129\,\text{mT}$ and $-0.011\%$ for
$B_\text{hold} = 120\,\text{mT}$. Both values are well below the
assumed precision ${\cal O}(0.1\%)$ of the jet-target polarization
determination using the Breit--Rabi polarimeter.

\begin{figure}[t]
\begin{center}
\includegraphics[width=0.95\columnwidth]{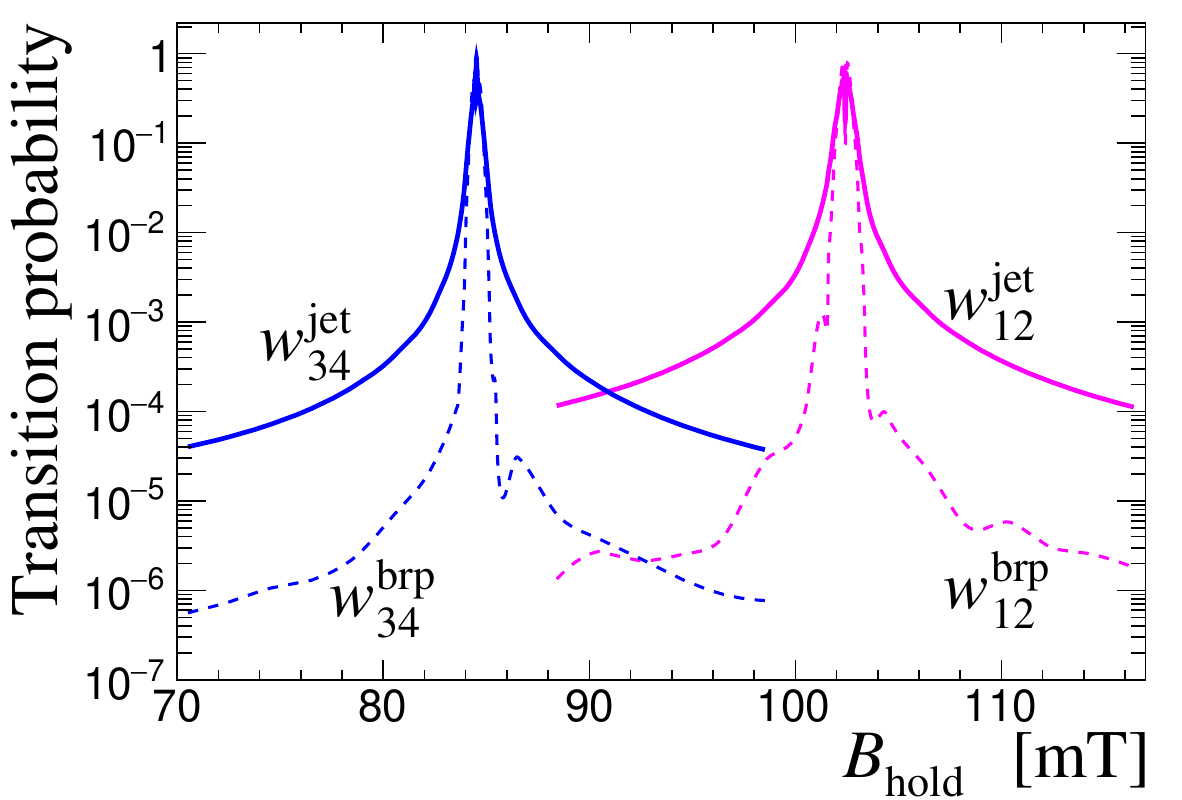}
\end{center}
\caption{\label{fig:12-34_res}
  Calculated transition probabilities $w^\text{jet}$ and $w^\text{brp}$ for the
  $f_{12}^\pi$ and $f_{34}^\pi$ transitions in the vicinity of the corresponding
  resonant values of the holding magnetic field.
}
\end{figure}

For completeness, we also evaluate the potential depolarization in the vicinity
of the resonant holding-field values corresponding to the $f_{12}^\pi$ and
$f_{34}^\pi$ transitions. The calculated transition probabilities
$w^\text{brp}$ and $w^\text{jet}$ for both transitions are shown in
Fig.\,\ref{fig:12-34_res}.

For $B_\text{hold}\!\approx\!120.45\,\text{mT}$, the transition probability
reaches its maximal possible value, $w_{12}^\text{jet}=1$, corresponding to a
complete spin flip of an atom initially in the state $|1\rangle$, i.e., full
conversion $|1\rangle\!\to\!|2\rangle$. Consequently,
this would imply complete depolarization of the jet target if it were formed by atoms in
states $|1\rangle$ and $|4\rangle$ corresponding to polarization $P_+$.

In contrast to the \emph{non-resonant} holding-field range, the BRP-related
probability $w_{12}^\text{brp}$ may also become large (up to unity) in the
resonant case. However, since the states $|1\rangle$ and $|2\rangle$ have the
same electron spin projection $+1/2$, such depolarization cannot be detected by
the Breit--Rabi polarimeter unless the second RF unit --- which was not employed
during RHIC HJET operation --- is used.

A similar conclusion applies to the $f_{34}^\pi$ transition at
$B_\text{hold}\!\approx\!84.52\,\text{mT}$. In this case, however, the resulting
depolarization cannot be detected by the BRP even if the second RF unit were to
be used in the measurement.

\section{Stability of the evaluated depolarization against variations of the proton beam parameters  \label{sec:Stability} }

\subsection{Reduced temporal bunch length}

As shown above, the strong suppression of resonant beam-induced depolarization
is primarily due to the smallness of the exponential factor entering the Rabi
frequency,
\begin{equation}
  \omega_R \propto
  \exp\!\left(-2\pi^2 f_{ij}^2 \sigma_t^2\right).
  \label{eq:fRexp}
\end{equation}
Therefore, for a fixed holding field $B_\text{hold}$ (and hence a fixed transition
frequency $f_{ij}$), a reduction of the rms temporal bunch length $\sigma_t$
may substantially increase the jet-target depolarization.

\begin{figure}[t]
\begin{center}
\includegraphics[width=0.95\columnwidth]{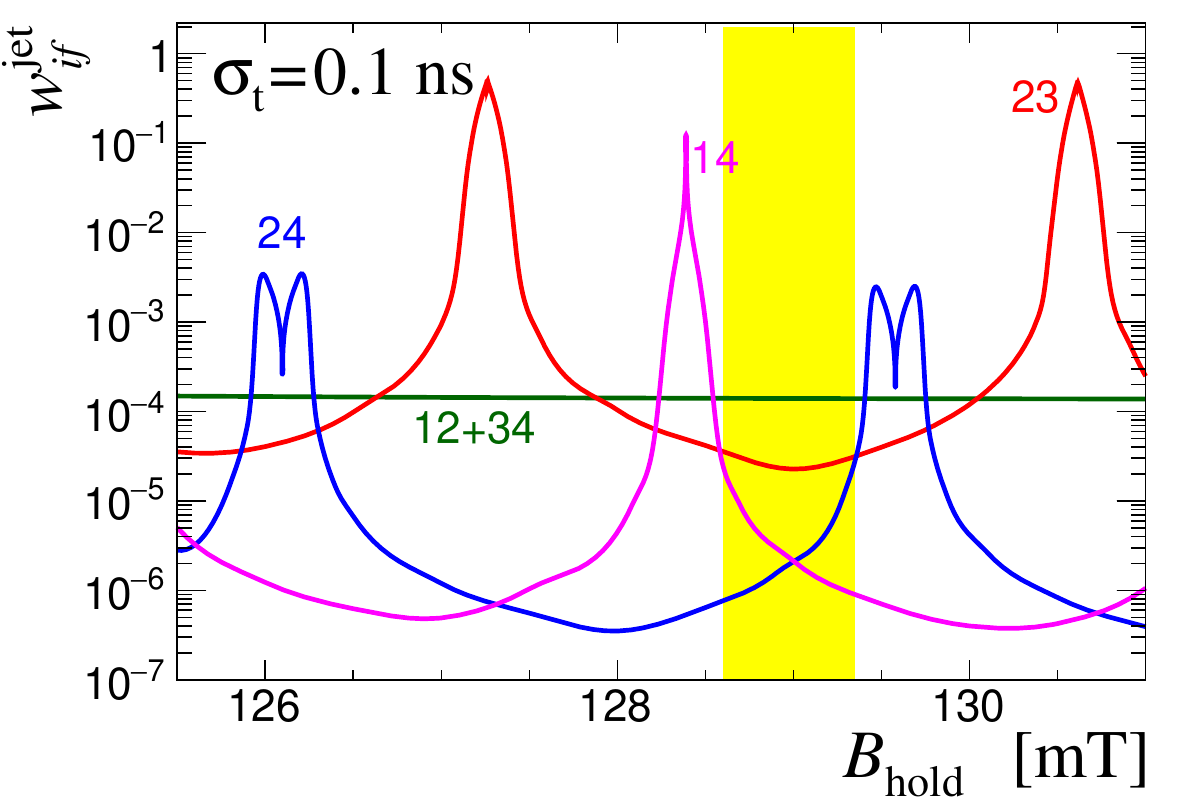}
\end{center}
\caption{\label{fig:st=0.1}
  Transition probabilities as functions of the holding magnetic field
for $\sigma_t = 0.1\,\text{ns}$. The shaded region indicates the holding-field
range $128.6 < B_\text{hold} < 129.35\,\text{mT}$, within which the overall
depolarization is negligibly small.}
\end{figure}

To estimate this effect, the dependence of the transition probabilities on
$B_\text{hold}$ was evaluated for all considered transitions assuming
$\sigma_t = 0.1\,\text{ns}$, which is a factor of two smaller than the nominal
EIC value. The results are shown in Fig.~\ref{fig:st=0.1}.

For a resonant transition, the probability \eqref{eq:Prob} can be written as
\begin{equation}
  w_\textit{res}^\text{jet}
  = \sin^2\!\left(\frac{\omega_R t_\text{int}}{2}\right)
  = \frac{1}{2}
    - \frac{1}{2}\cos\!\left(\omega_R t_\text{int}\right).
\end{equation}
In the regime $\omega_R t_\text{int} \gg 1$, and after averaging over a large
ensemble of hydrogen atoms, the oscillatory term is effectively canceled due to
variations in $v_\text{at}$, $x(0)$, and the relative phase between atomic
trajectories and the bunch structure.

Consequently, the limiting value $w_\textit{res}^\text{jet} = 0.5$ observed for the
$f_{23}^\pi$ transition in Fig.~\ref{fig:st=0.1} corresponds to complete
saturation of the transition, i.e.\ to equal populations of the initial state
$|2\rangle$ and the final state $|3\rangle$.

For the $\sigma$ transition $f_{24}^\sigma$, both $\omega_R$ and $\Delta\omega$
are primarily determined by the vertical component of the beam-induced magnetic
field $B_\text{ind}^{(y)}(r,t)$.
Since $B_\text{ind}^{(y)}$ changes sign for $x(0) < 0$ and $x(0) > 0$, the
resonance condition is satisfied at two slightly different values of
$B_\text{hold}$, resulting in a characteristic double-peak structure clearly
visible in Fig.~\ref{fig:st=0.1}.

Although the depolarization probabilities for $\sigma_t = 0.1\,\text{ns}$ are
significantly larger than those obtained for the nominal
$\sigma_t = 0.2\,\text{ns}$, depolarization-free operation of the HJET remains
possible provided that the magnetic field in the scattering chamber is
maintained within
\begin{equation}
  B_\text{hold} = 129 \pm 0.35\,\text{mT},
\end{equation}
prior to beam-induced corrections.

A further reduction of $\sigma_t$ would require a corresponding increase of the
holding field (and hence of the transition frequencies $f_{ij}$) in order to
suppress resonant depolarization.
Such a scenario is undesirable, since increasing $B_\text{hold}$ would lead to
a noticeable increase of systematic uncertainties in proton beam
polarization measurements performed with the HJET recoil spectrometer.

\subsection{Increased beam current}

The dependence of the \emph{non-resonant} depolarization probability on the beam
parameters can be approximately expressed as
\begin{equation}
  w_\text{non-res}^\text{jet}
  \propto B_\text{ind}^{\text{max}\,2}
  \propto \left(\frac{I_\text{avg}}{\sigma_r}\right)^2,
  \label{eq:I/sr}
\end{equation}
where $I_\text{avg}$ is the average beam current and $\sigma_r$ is the rms beam
radius.

Thus, increasing $I_\text{avg}$ (or decreasing $\sigma_r$) by a factor of five
is expected to increase the depolarization probability by approximately a
factor of $25$.
To verify this scaling, dedicated calculations were performed, and the results
are summarized in Table~\ref{tab:I/sr}.

\begin{table}[t]
  \caption{\label{tab:I/sr}
    Dependence of the \emph{non-resonant} depolarization probability on the beam
    current and transverse beam size for $B_\text{hold}=129\,\text{mT}$.
  }
  \begin{center}
    \begin{tabular}{c c | c c}
      $I_\text{avg}$ [A] & $\sigma_r$ [mm] &
      $B_\text{pk}^\text{max}$ [mT] & $w_\text{non-res}^\text{jet}$ \\[3pt]
      \hline \\[-7pt]
      1 & 0.656 &  3  & $7.82\times10^{-5}$ \\
      5 & 0.656 & 15  & $1.95\times10^{-3}$ \\
      1 & 0.131 & 15  & $1.94\times10^{-3}$
    \end{tabular}
  \end{center}
\end{table}

The results confirm that the scaling~\eqref{eq:I/sr} provides a reliable estimate.
Even for a fivefold increase of the average beam current (assuming unchanged
$\sigma_t$ and $\sigma_r$), the resulting depolarization of
$\sim 2\times10^{-3}$ remains fairly comfortable within the EIC polarization accuracy
requirement~\eqref{eq:EICreq}.

\subsection{Elliptical beam profile}

The planned 255~GeV EIC proton beam has an elliptical transverse profile,
\begin{equation}
  I(x,y) \propto
  \frac{1}{2\pi\sigma_x\sigma_y}
  \exp\!\left(-\frac{x^2}{2\sigma_x^2}
             -\frac{y^2}{2\sigma_y^2}\right),
\end{equation}
with $\sigma_x=1.610\,\text{mm}$ and $\sigma_y=0.268\,\text{mm}$.

For simplicity, the calculations presented above assumed a round beam with rms
radius $\sigma_r=(\sigma_x\sigma_y)^{1/2}$, which could potentially distort the
estimate of the beam-induced depolarization.

As it follows from the Amp\`ere--Maxwell law \eqref{eq:AM},
for a strongly elongated beam ($\sigma_y\ll\sigma_x$), the induced magnetic
field near the jet is systematically smaller than that produced by a round
beam with the same rms radius $\sigma_r$.
Therefore, the actual depolarization is expected to be lower than that
evaluated assuming a round beam.

To quantify this effect, the beam-induced magnetic field
$\boldsymbol{B}_\text{pk}(x,y)$ for an elliptical current distribution was
calculated numerically and used in the hydrogen-atom tracking.
For the \emph{non-resonant} transitions and $B_\text{hold}=129\,\text{mT}$, the
resulting depolarization was found to be
\begin{equation}
  w_\text{non-res}^\text{jet} = 4.76\times10^{-5},
\end{equation}
which is a factor of 1.6 lower than the corresponding value in
Table~\ref{tab:I/sr} obtained within the round-beam approximation.

\subsection{Depolarization rate dependence on the holding field value}

The depolarization rate exhibits a dual dependence on the value of the holding field.

\noindent{---}\quad
For the transitions
$|1\rangle\!\leftrightarrow\!|2\rangle$,
$|3\rangle\!\leftrightarrow\!|4\rangle$, and
$|2\rangle\!\leftrightarrow\!|4\rangle$,
the matrix elements \eqref{eq:Msigma} and \eqref{eq:Mpi}, and consequently the Rabi frequencies, depend on the mixing angle $\theta$ through a factor $\sin\theta$. Since $\theta=\pi/4$ for $B_\text{hold}=0$, a decrease of the holding field may increase the depolarization rate for these transitions by up to an order of magnitude.

\noindent{---}\quad
For the \emph{potentially resonant} transitions, the transition frequency $f_{ij}$ depends strongly on the holding field (see Fig.\,\ref{fig:HFS}). Since the corresponding Rabi frequency depends exponentially on $f_{ij}$ [see Eq.\,\eqref{eq:fRexp}], the depolarization rate may increase dramatically as $B_\text{hold}$ is reduced. As a simple estimate of this effect, one may interpret the transition probabilities shown in Fig.\,\ref{fig:st=0.1} as corresponding to $\sigma_t=0.2\,\text{ns}$ but for an \emph{approximately} factor-of-two smaller holding field.

This dependence of the beam-induced depolarization on the holding field suggests a possible method for experimental evaluation and normalization of the effect. However, a detailed investigation of this possibility is beyond the scope of the present paper.

\section{Comparison of Beam-Induced Depolarization Across Different Experimental Conditions \label{sec:Benchmark}}

\subsection{EIC injection}

\begin{figure}[t]
  \begin{center}
    \includegraphics[width=0.95\columnwidth]{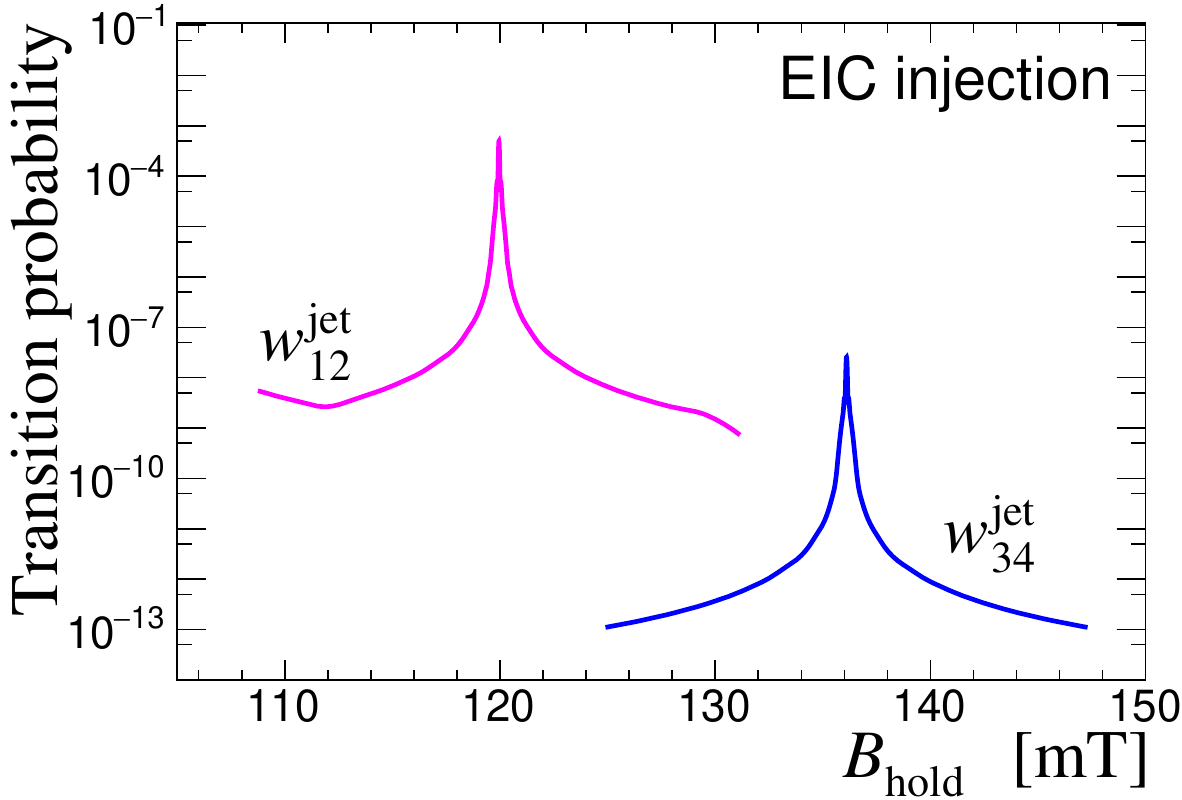}
  \end{center}
  \caption{\label{fig:12-34_inj}
    Calculated transition probabilities $w^\text{jet}$ for the
    $f_{12}^\pi$ and $f_{34}^\pi$ transitions in the vicinity of the corresponding
    resonant values of the holding magnetic field for the EIC injection.
  }
\end{figure}

Compared to the EIC flattop, the EIC injection beam has a factor of 2.4 larger transverse size and a factor of 4 longer bunch length. This results in an approximate reduction of the resonant harmonic amplitude by a factor of 160 for the $|1\rangle\!\leftrightarrow\!|2\rangle$ transition, and in a much stronger suppression for the other transitions. The transition probability, as a function of the holding field, is shown in Fig.\,\ref{fig:12-34_inj}. For $B_\text{hold}=129\,\text{mT}$, it is below $10^{-8}$.

\subsection{RHIC flattop}

At RHIC, the bunch length was a factor of 9.2 larger than that planned for the EIC flattop. Consequently, in the vicinity of $B_\text{hold}=120\,\text{mT}$, the resonant harmonic amplitudes are strongly suppressed (by at least a factor of $3\times10^9$ compared to the EIC flattop), and the corresponding depolarization transition probability is negligible, below $10^{-20}$.

\subsection{Beam-induced depolarization in the HERMES experiment \label{sec:HERMES}}

The experimental observation of beam-induced depolarization due to the resonance transitions $|1\rangle\to|2\rangle$ and $|4\rangle\to|3\rangle$ in hydrogen atoms was reported by the HERMES Collaboration \cite{HERMES:1998twm}. In these measurements, an electron beam with the following parameters was used:
$I_\text{avg}=0.045\,\text{A}$,
$\sigma_r=\sqrt{\sigma_x\sigma_y}=0.15\,\text{mm}$,
$\sigma_t=27\,\text{ps}$,
$f_b=10.666\,\text{MHz}$.

The holding field $B_\text{hold}$ in the range 220--400\,mT was directed along the beam ($z$-axis). A detailed study of the depolarization was performed for $B_\text{hold}\approx366\,\text{mT}$. Within the $\pm200\,\text{mm}$ interaction region, the value of $B_\text{hold}$ varied within a $\pm5\,\text{mT}$ range, which effectively broadened the resonance intensity distribution as a function of the average $B_\text{hold}$. For the estimates presented here, this $z$-dependence of the holding field is neglected to simplify the analysis.

Comparing the experimental conditions at HERMES with those at the EIC HJET, we note that the beam-induced magnetic field at HERMES was always orthogonal to the holding field. Consequently, there were no beam-induced variations of $B_\text{hold}$ and no $|2\rangle\leftrightarrow|4\rangle$ resonance transitions. In addition, for $B_\text{hold}\approx366\,\text{mT}$, the $|4\rangle\to|3\rangle$ transition is strongly suppressed. Therefore, only the $|1\rangle\!\leftrightarrow\!|2\rangle$, $|1\rangle\!\leftrightarrow\!|4\rangle$, and $|2\rangle\!\leftrightarrow\!|3\rangle$ resonances need to be considered.

The approximately three times larger holding field at HERMES enhances the exponential suppression of the harmonic amplitudes. However, this effect is more than compensated by the factor of $\sim 8$ shorter bunch length $\sigma_t$ at HERMES.

Additional factors that must be taken into account include the reduced depolarization due to the lower beam current $I_\text{avg}$ and the further suppression of the matrix elements by the holding-field-dependent factor $\sin\theta$ at HERMES, as well as enhancing effects such as the smaller beam radius $\sigma_r$.

\begin{table}[t]
  \caption{\label{tab:HERMES}
    Evaluation of the resonant transition probabilities for the HERMES beam and holding field. Calculations are performed separately for the HJET-like target approximation and for the HERMES storage cell.
  }
  \begin{center}
    \begin{tabular}{l | c | c}
      & HJET & HERMES \\
      & $w$ $(v_\text{at}=1.8\,\text{mm/}\mathrm{\mu}\text{s})$
      & $w_\text{avg}$ (cell target) \\
      \hline & & \\[-7pt]
      $f_{12}^\pi$  & $1.3\times10^{-4}$ & 0.15 \\
      $f_{14}^\pi$  & $8.7\times10^{-4}$ & 0.23 \\
      $f_{23}^\pi$  & $2.0\times10^{-3}$ & 0.27
    \end{tabular}
  \end{center}
\end{table}

Assuming that the polarized hydrogen target consists of vertically moving atoms with velocity $v_\text{at}=1800\,\text{m/s}$ (as at HJET), the transition probabilities $w$ were calculated within a two-level approximation for the HERMES beam and holding field. The results are presented in Table~\ref{tab:HERMES}.

Notably, the $|1\rangle\!\leftrightarrow\!|4\rangle$ and $|2\rangle\!\leftrightarrow\!|3\rangle$ transition probabilities are significantly larger than those evaluated for the EIC beam and the RHIC HJET holding field. However, the resulting beam depolarization inferred from these $w$ values remains too small to explain the HERMES observations.

It should be noted that the calculated $w$ is inversely proportional to the square of the atomic velocity $v_\text{at}$. Consequently, the low transverse velocity component $v_\perp$ of the polarized target atoms may significantly enhance the depolarization probability. At HERMES, a storage cell target cooled to $T=100\,\text{K}$ was used. According to the Maxwell--Boltzmann distribution,
\begin{equation}
  \frac{dN}{dv_\perp}\propto f(v_\perp)=v_\perp
  \exp\!\left(-\frac{mv_\perp^2}{2k_BT}\right).
\end{equation}

When evaluating the average depolarization, one must also account for the fact that the interaction probability of a target atom with the beam is inversely proportional to $v_\text{at}$. Therefore, the average resonance transition probability at HERMES can be estimated as
\begin{equation}
  w_\text{avg} =
  \frac{\int dv_\perp\,v_\perp^{-1} f(v_\perp)
    \sin^2\!\left( \sqrt{w}\,v_\text{at}/v_\perp \right)}
       {\int dv_\perp\,v_\perp^{-1} f(v_\perp)}.
       \label{eq:cell}
\end{equation}
The results of these calculations are also shown in Table~\ref{tab:HERMES}.

The calculated depolarization of about 15\% is in reasonably good agreement with the peak value of the measured depolarization at HERMES (see Fig.\,5 of Ref.\,\cite{HERMES:1998twm}). Nevertheless, it should be emphasized that the present calculation is based on an oversimplified description of the HERMES experiment.

For example, possible effects of concurrent $|1\rangle\!\leftrightarrow\!|4\rangle$ and $|2\rangle\!\leftrightarrow\!|3\rangle$ transitions were not included in the estimate. However, these transitions have only a minor impact on the nuclear polarization of the hydrogen atom, and their overall effect is expected to be small.

In addition, variations of the holding field along the beam direction were not considered. Nevertheless, the calculated depolarization depends critically on the low-velocity part of the transverse velocity distribution, where the depolarization effect is effectively saturated by the $\sin^2\!\left( \sqrt{w}\,v_\text{at}/v_\perp \right)$ factor. Therefore, the reduction of depolarization due to non-resonant (but small detuning) values of the holding field may be partially mitigated. For example, although the $|2\rangle\!\leftrightarrow\!|3\rangle$ transition probability $w$ is about 15 times larger than that for $|1\rangle\!\leftrightarrow\!|2\rangle$, the ratio for $w_\text{avg}$ is only about 1.8.

For completeness, it should be noted that in Ref.\,\cite{HERMES:1998twm}, the analysis was based on calculations using an \emph{average} transition probability, which implicitly corresponds to the low-velocity approximation. The predictions of that model were normalized to the data by an arbitrary factor. In this context, the estimate of the depolarization presented here may be regarded a refining of the model used in Ref.\,\cite{HERMES:1998twm}.

\subsection{Beam-induced depolarization at LHCspin}

The beam-induced depolarization in polarized gas targets for the LHCspin~\cite{LHCspin:2025lvj} project was evaluated in Refs.~\cite{Steffens:2018gzz,Lenisa:2020bxj}. The estimate was based on comparing the Rabi frequency $\omega_R$ of the
$|2\rangle\!\leftrightarrow\!|4\rangle$ transition at LHCspin and in the HERMES experiment, assuming the proportionality
\begin{equation}
\omega_R \propto
I_\text{avg}
\exp{\left(
-2\pi^2 f_{24}^2 \sigma_t^2
\right)}.
\label{eq:omegaLHC}
\end{equation}
Although
$I_\text{avg}^\text{LHC}/I_\text{avg}^\text{HERMES}\approx25$,
the substantially larger bunch length at the LHC,
$\sigma_t^\text{LHC}/\sigma_t^\text{HERMES}\approx8.2$,
leads, for the considered transition frequency
$f_{24}=8.54\,\text{GHz}$, to
\begin{equation}
\omega_R^\text{LHC}/\omega_R^\text{HERMES}
= 3.8\times10^{-40}.
\end{equation}
Consequently, the beam-induced depolarization associated with the
$|2\rangle\!\leftrightarrow\!|4\rangle$
transition may be regarded as entirely negligible at LHCspin.

Notably, the proportionality~\eqref{eq:omegaLHC} employed in Refs.~\cite{Steffens:2018gzz,Lenisa:2020bxj} is fully consistent with the expressions for the Rabi frequency derived in the present work, as follows from Eqs.~\eqref{eq:omegaR_max}, \eqref{eq:zero}, and \eqref{eq:Bpk}, except that the rms beam radius $\sigma_r$ was not explicitly considered in Refs.~\cite{Steffens:2018gzz,Lenisa:2020bxj}. For a fixed beam current, the beam radius determines the maximum peak value of the induced magnetic field and is therefore, in general, an important parameter when comparing beam-induced depolarization effects among different experiments.

Assuming an rms beam radius
$\sigma_r=0.15\,\text{mm}$
(as in HERMES) and applying the calculation framework developed in the present paper, one obtains a
$|2\rangle\!\leftrightarrow\!|4\rangle$
transition probability of
$3\times10^{-83}$
at LHCspin for a
$300\,\text{mT}$
vertical holding field and an EIC-like jet target. Since a storage-cell target is planned for LHCspin, the corresponding beam-induced depolarization, corrected according to Eq.~\eqref{eq:cell}, can be estimated as
\begin{equation}
w_{24}^\text{LHC} \approx 2\times10^{-21}.
\end{equation}
Although smaller beam radii would lead to larger values of
$w_{24}^\text{LHC}$, the beam-induced depolarization remains negligible for any realistic value of $\sigma_r$.

\subsection{Alternative evaluation~\cite{Rathmann:2025jgp} of beam-induced depolarization at the EIC}

As already noted in the Introduction, Ref.~\cite{Rathmann:2025jgp} reported a very large expected depolarization for EIC HJET operation with a $120\,\text{mT}$ holding field. However, a detailed examination of the procedure used to evaluate the Rabi frequencies and, consequently, the resonance transition strengths indicates that the depolarization was substantially overestimated.

The relative harmonic amplitudes, both in Ref.~\cite{Rathmann:2025jgp} and in the present work, are modulated by the same Gaussian envelope. Using the harmonic-amplitude normalization~\eqref{eq:zero}, one finds for the average field
$\langle B(f)\rangle$
(defined in Eq.\,(37) of Ref.~\cite{Rathmann:2025jgp})
\begin{equation}
\langle B(0)\rangle = 25\,\mathrm{\mu T},
\qquad
\sum_f \langle B(f)\rangle = 550\,\mathrm{\mu T}.
\end{equation}
Here, the sum over all harmonic amplitudes is, by construction, equal to the peak value of the spatially averaged magnetic field.

In Ref.\,\cite{Rathmann:2025jgp}, no explanation of the harmonic amplitude normalization is provided, and the assigned values cannot be considered justified. For example, Fig.\,16 of Ref.~\cite{Rathmann:2025jgp} shows $\langle B(0)\rangle \approx 1500\,\mu\text{T}$. Moreover, if the reported 6th harmonic amplitude of $1174\,\mu\text{T}$ were correct, the sum of only the first seven harmonics, according to Eq.\,\eqref{eq:Bind}, would exceed $15.3\,\text{mT}$, which is already far larger than the peak field of $3.03\,\text{mT}$. Thus, the Rabi frequency was overestimated by approximately a factor of 60, and the depolarization probability by up to a factor of about 3600.

Furthermore, to evaluate the depolarization probability at the EIC flattop due to the $|2\rangle\!\leftrightarrow\!|4\rangle$ transition, the following well-known formula,
\begin{equation}
 \frac{\omega_R^2}{\Omega^2}\sin^2\!\left(\frac{\Omega t}{2}\right)
 \quad\xrightarrow{\Delta\omega=0}\quad
 \sin^2\!\left(\frac{\omega_R t}{2}\right),
 \label{eq:Rabi}
\end{equation}
(which can be readily derived from Eq.\,\eqref{eq:Prob}), was used in Ref.\,\cite{Rathmann:2025jgp} under the assumption $\Delta\omega=0$. To estimate the corresponding Rabi frequency, an effective field amplitude of $2.0\,\text{mT}$ --- of the order of the peak value $B_\text{pk}^\text{max}=3\,\text{mT}$ --- was inferred using a ``spatial field distribution'' approach. However, this analysis does not take into account that the variation frequency of the spatial magnetic field is fixed by the bunch frequency $f_b$. Consequently, a large detuning $\Delta\omega$ must be included in Eq.\,\eqref{eq:Rabi}, which strongly suppresses the transition probability.

Due to the superposition principle, a resonant transition can be driven only by the corresponding resonant harmonic in the Fourier expansion of the spatial magnetic field. Therefore, the Gaussian spectral envelope suppression of the resonant amplitude must be properly taken into account~\cite{Steffens:2018gzz,Lenisa:2020bxj}. For the EIC flattop beam, the suppression of the resonant $|2\rangle\!\leftrightarrow\!|4\rangle$ harmonic is approximately $7\times10^{-6}$ (see Table~\ref{tab:EIC-129}), which effectively reduces the corresponding beam-induced depolarization probability to a negligible level.

Thus, even without addressing other aspects of Ref.\,\cite{Rathmann:2025jgp}, correcting the harmonic amplitudes to physically consistent values alone leads to the conclusion that Ref.\,\cite{Rathmann:2025jgp} does not provide evidence that beam-induced depolarization of the HJET target is significant under the considered EIC flattop beam conditions.

It is also worth recalling (see Section~\ref{sec:PotRes}) that an upper limit for the resonant transition probability can be readily obtained by calculating the corresponding Rabi frequency for the peak value of the induced magnetic field~\eqref{eq:omegaR_max} and assuming exactly zero detuning over the entire time the atom traverses the scattering chamber. In this extreme case, the $|2\rangle\!\leftrightarrow\!|4\rangle$ transition probability can be evaluated in a straightforward and reproducible manner. The resulting value, $8\times10^{-7}$~\eqref{eq:upper24}, clearly excludes the much larger $|2\rangle\!\leftrightarrow\!|4\rangle$ depolarization probability reported in Ref.\,\cite{Rathmann:2025jgp}.

\subsection{$\sigma$ transition in an SFT unit}

Equations~\eqref{eq:an(t)} utilized in this work to evaluate beam-induced depolarization were also employed in Refs.~\cite{Beijers:2005,Schieck:2008} to describe the evolution of hydrogen atoms in WFT and SFT units. Within the framework adopted here, the SFT transition $|2\rangle \to |4\rangle$ can be readily modeled by replacing
\begin{align}
  B_\text{hold}^\text{eff}(r,t) &\to B_\text{hold}(y)=B_0+B_yy=\left(10+y/\text{cm}\right)\,\text{mT},
  \\
  \omega_\eta/2\pi &\to \omega_\text{SFT}/2\pi = 1.448\,\text{GHz},
  \\
  \omega_R(t) &\to \omega_R^\sigma = \mu_{24}^\sigma B_\text{RF}/2\hbar,
\end{align}
in Eqs.~\eqref{eq:omegaR} and \eqref{eq:omegaD}.

\begin{figure}[t]
  \begin{center}
    \includegraphics[width=0.95\columnwidth]{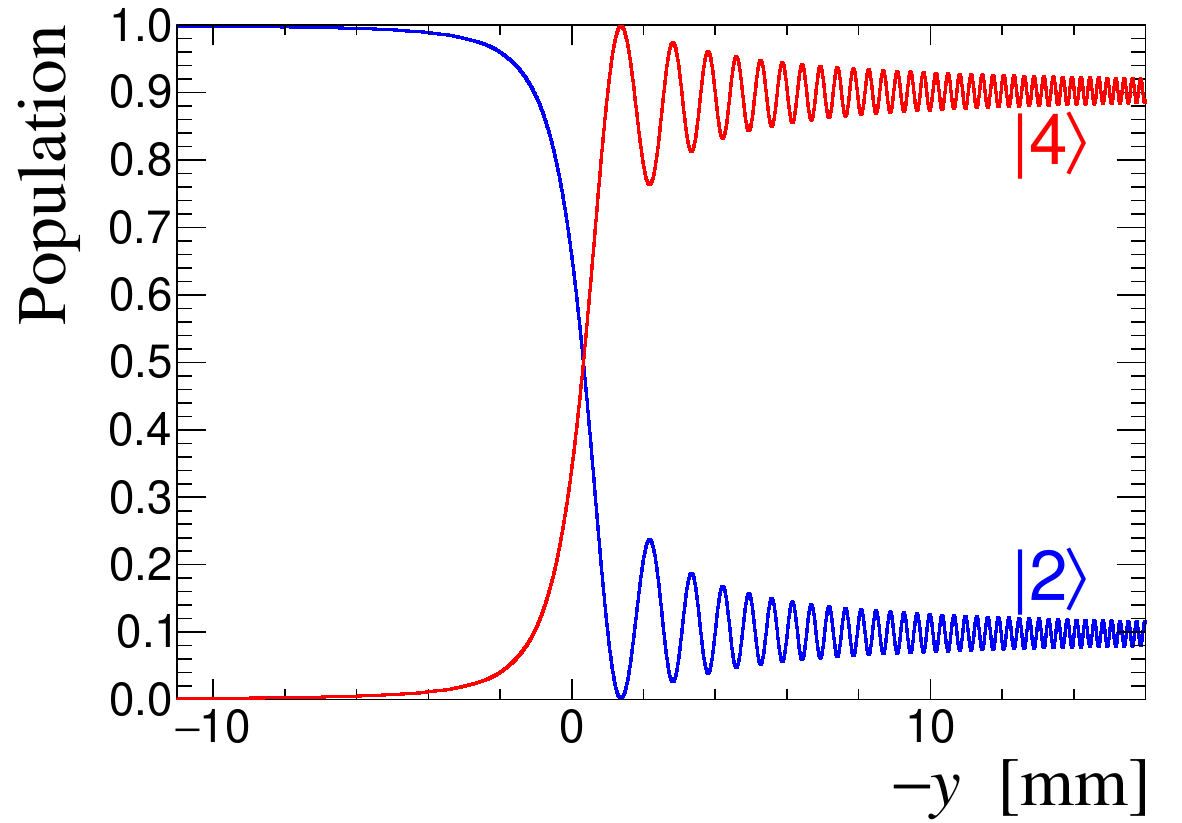}
  \end{center}
  \caption{\label{fig:SFT}
    Calculated evolution of the $|2\rangle$ and $|4\rangle$ state populations along the atomic trajectory in the SFT unit. The atom is initially in state $|2\rangle$.
  }
\end{figure}

To illustrate the applicability of the present approach to adiabatic RF transitions, the evolution of the hydrogen atom levels was calculated for a relatively weak RF field, $B_\text{RF}=0.35\,\text{G}$, resulting in a transition inefficiency of approximately
\begin{equation}
  \tilde{\epsilon}_{24}^\text{calc}\approx9.76\%.
\end{equation}
The calculated variation of the level populations along the atomic trajectory, shown in Fig.~\ref{fig:SFT}, is structurally similar to that presented in Fig.~1(a) of Ref.~\cite{Beijers:2005}. For the SFT unit parameters used, the analytical approach~\cite{Carroll:1986} yields the following estimate, in good agreement with $\tilde{\epsilon}_{24}^\text{calc}$:
\begin{equation}
  \tilde{\epsilon}_{24}^\text{theor} = \exp{\left(%
    -\frac{\pi^2}{4} \frac{f_\text{hft}}{v_\text{at}}%
       \frac{B_\text{RF}^2}{B_0B_y} \sin{2\theta_0}
      \right)}%
    =9.64\% ,
\end{equation}
where $\theta_0$ is the mixing angle corresponding to the holding field $B_0$. Practically the same expression for $\tilde{\epsilon}_{24}^\text{theor}$ was presented in Ref.~\cite{Beijers:2005}, with reference to the analytical solution of the time-dependent Schr\"odinger equation for a two-level Hamiltonian by Landau~\cite{Landau:1932vnv} and Zener~\cite{Zener:1932ws}.

Increasing the RF field to $0.7\,\text{G}$ results in a smoother evolution of the populations along the atomic trajectory and yields a calculated transition efficiency exceeding 99.9\%.

\section{Summary}

In this work, the possibility of beam-induced depolarization of the HJET target
in the EIC environment has been evaluated.

As a prerequisite for this study, the methods used to determine the jet target
polarization with the Breit--Rabi polarimeter and to apply this
polarization in the recoil spectrometer were compared. It was noted that, for
the determination of the proton beam polarization with the recoil spectrometer,
only the average magnitude of the jet spin-up and spin-down polarizations,
$|P_\pm|$, is required. Consequently, Eq.\,\eqref{eq:P_BRP} can be used for a
precise evaluation of the jet target polarization $P_\text{jet}$. Since the
ratio~\eqref{eq:m0/m+-} is very small, even relative systematic uncertainties of
order 50\% in its determination can be regarded as negligible.

To evaluate beam-induced depolarization of the jet target, atomic hydrogen was treated as a four-level hyperfine system in an external (holding) magnetic field, perturbed by a time-dependent magnetic field generated by the bunched proton beam. Two main effects were considered: transitions between hyperfine levels induced by a high-frequency beam harmonic resonant with a given transition frequency, and an effective modification of the holding field (and, consequently, of the atomic energy levels) caused by the coherent contribution of low-frequency harmonics.

First, it was demonstrated that resonant conditions that may occur during the
tracking of hydrogen atoms through the scattering chamber do not lead to
significant transitions, because the corresponding high-frequency harmonics do
not carry sufficient power for the nominal EIC flattop beam parameters.

Under these conditions, the four-level hydrogen system naturally separates into
two independent two-level subsystems, $|1\rangle$--$|2\rangle$ and
$|3\rangle$--$|4\rangle$. Although the beam-induced field harmonics driving
these transitions are substantially stronger, the corresponding resonance
conditions can be readily avoided by an appropriate choice of the holding field
$B_\text{hold}$.

The evolution of these subsystems during atomic traversal through the proton beam was evaluated by numerically solving Eqs.\,\eqref{eq:alpha}, with time-dependent Rabi frequencies $\omega_R(t)$ and detunings $\Delta\omega(t)$. Notably, the approach used is also applicable to approximating adiabatic SFT transitions.

As a result, for the nominal EIC proton beam at flattop and for the RHIC operating holding field $B_\text{hold}\sim120\,\text{mT}$, if applied to the EIC HJET, the expected beam-induced depolarization of the jet target,
\begin{equation}
  |\Delta_\text{dep} P_\text{jet}| \lesssim 0.01\%,
\end{equation}
is found to be negligible.

Possible variations of the EIC flattop beam parameters that preserve this low level of depolarization were also examined.

Importantly, the method used here to evaluate beam-induced depolarization is based directly on the fundamental quantum-mechanical equations, namely the Schr"odinger equation for a time-dependent Hamiltonian. Since the hyperfine Hamiltonian for the hydrogen atom in the ground state is well established, the approach may be regarded as a first-principles calculation. Consequently, the accuracy of the depolarization estimate is primarily limited by the numerical integration of the quantum-mechanical evolution equations and by possible uncertainties in the description of the holding field and proton beam parameters.

In fact, essentially the same approach has previously been employed to calculate transition probabilities in SFT and WFT units~\cite{Beijers:2005,Schieck:2008}. Since the calculation framework developed in the present paper can be readily adapted to studies of SFT transitions, such a calculation was also performed. The resulting SFT transition efficiency was found to be in good agreement with the corresponding theoretical prediction.

\section*{Acknowledgments}
The author was motivated to write this paper through numerous discussions with
Frank Rathmann, Alexander Nass, and Oleg Eyser.
This manuscript was authored by an employee of Brookhaven Science Associates,
LLC, under Contract No.~DE-SC0012704 with the U.S. Department of Energy.

\bibliographystyle{elsarticle-num}
%\bibliography{v1.4_BeamInducedDepolarization.bib}

\begin{thebibliography}{10}
\expandafter\ifx\csname url\endcsname\relax
  \def\url#1{\texttt{#1}}\fi
\expandafter\ifx\csname urlprefix\endcsname\relax\def\urlprefix{URL }\fi
\expandafter\ifx\csname href\endcsname\relax
  \def\href#1#2{#2} \def\path#1{#1}\fi

\bibitem{Accardi:2012qut}
A.~Accardi, J.~L. Albacete, M.~Anselmino, N.~Armesto, E.-C. Aschenauer,
  A.~Bacchetta, D.~Boer, W.~K. Brooks, T.~Burton, N.-B. Chang, et~al.,
  {Electron Ion Collider: The Next QCD Frontier}, Eur. Phys. J. A 52~(9) (2016)
  268.
\newblock \href {http://arxiv.org/abs/1212.1701} {\path{arXiv:1212.1701}},
  \href {https://doi.org/10.1140/epja/i2016-16268-9}
  {\path{doi:10.1140/epja/i2016-16268-9}}.

\bibitem{Willeke:2021ymc}
F.~Willeke, {Electron Ion Collider Conceptual Design Report 2021}, Tech. Rep.
  BNL-221006-2021-FORE, Brookhaven National Laboratory, Upton NY (2 2021).
\newblock \href {https://doi.org/10.2172/1765663} {\path{doi:10.2172/1765663}}.

\bibitem{AbdulKhalek:2021gbh}
R.~Abdul~Khalek, A.~Accardi, J.~Adam, D.~Adamiak, W.~Akers, M.~Albaladejo,
  A.~Al-bataineh, M.~G. Alexeev, F.~Ameli, P.~Antonioli, et~al., {Science
  Requirements and Detector Concepts for the Electron-Ion Collider}: {EIC
  Yellow Report}, Nucl. Phys. A 1026 (2022) 122447.
\newblock \href {http://arxiv.org/abs/2103.05419} {\path{arXiv:2103.05419}},
  \href {https://doi.org/10.1016/j.nuclphysa.2022.122447}
  {\path{doi:10.1016/j.nuclphysa.2022.122447}}.

\bibitem{Bunce:2000uv}
G.~Bunce, N.~Saito, J.~Soffer, W.~Vogelsang, {Prospects for spin physics at
  RHIC}, Ann. Rev. Nucl. Part. Sci. 50 (2000) 525--575.
\newblock \href {http://arxiv.org/abs/hep-ph/0007218}
  {\path{arXiv:hep-ph/0007218}}, \href
  {https://doi.org/10.1146/annurev.nucl.50.1.525}
  {\path{doi:10.1146/annurev.nucl.50.1.525}}.

\bibitem{Aschenauer:2015eha}
E.-C. Aschenauer, A.~Bazilevsky, M.~Diehl, J.~Drachenberg, K.~O. Eyser,
  R.~Fatemi, C.~Gagliardi, Z.~Kang, Y.~V. Kovchegov, J.~Lajoie, et~al., {The
  RHIC SPIN Program: Achievements and Future Opportunities} (1 2015).
\newblock \href {http://arxiv.org/abs/1501.01220} {\path{arXiv:1501.01220}}.

\bibitem{Zelenski:2013zxa}
A.~N. Zelenski, G.~Atoian, A.~A. Bogdanov, S.~B. Nurushev, F.~S. Pylaev,
  D.~Raparia, M.~F. Runtso, E.~Stephenson, {Absolute polarimeter for the
  proton-beam energy of 200 MeV}, Phys. Atom. Nucl. 76 (2013) 1490--1496.
\newblock \href {https://doi.org/10.1134/S1063778813120156}
  {\path{doi:10.1134/S1063778813120156}}.

\bibitem{Poblaguev:2025ZO}
A.~Poblaguev, G.~Atoian, A.~Cannavo, A.~Zelenski, {New DAQ System for the 200
  MeV Polarimeter at BNL Linac}, PoS PSTP2024 (2025) 009.
\newblock \href {https://doi.org/10.22323/1.472.0009}
  {\path{doi:10.22323/1.472.0009}}.

\bibitem{Huang:2006cs}
H.~Huang, K.~Kurita, {Fiddling carbon strings with polarized proton beams}, AIP
  Conf. Proc. {868} (2006) 3--21.
\newblock \href {https://doi.org/10.1063/1.2401392}
  {\path{doi:10.1063/1.2401392}}.

\bibitem{Zelenski:2005mz}
A.~Zelenski, A.~Bravar, D.~Graham, W.~Haeberli, S.~Kokhanovski, Y.~Makdisi,
  G.~Mahler, A.~Nass, J.~Ritter, T.~Wise, V.~Zubets, {Absolute polarized H-jet
  polarimeter development, for RHIC}, Nucl. Instrum. Meth. A 536 (2005)
  248--254.
\newblock \href {https://doi.org/10.1016/j.nima.2004.08.080}
  {\path{doi:10.1016/j.nima.2004.08.080}}.

\bibitem{Poblaguev:2020qbw}
A.~A. Poblaguev, A.~Zelenski, G.~Atoian, Y.~Makdisi, J.~Ritter, {Systematic
  error analysis in the absolute hydrogen gas jet polarimeter at RHIC}, Nucl.
  Instrum. Meth. A 976 (2020) 164261.
\newblock \href {http://arxiv.org/abs/2006.08393} {\path{arXiv:2006.08393}},
  \href {https://doi.org/10.1016/j.nima.2020.164261}
  {\path{doi:10.1016/j.nima.2020.164261}}.

\bibitem{Poblaguev:2019saw}
A.~A. Poblaguev, A.~Zelenski, E.~Aschenauer, G.~Atoian, K.~O. Eyser, H.~Huang,
  Y.~Makdisi, W.~B. Schmidke, I.~Alekseev, D.~Svirida, N.~H. Buttimore,
  {Precision Small Scattering Angle Measurements of Elastic Proton-Proton
  Single and Double Spin Analyzing Powers at the RHIC Hydrogen Jet
  Polarimeter}, Phys. Rev. Lett. 123~(16) (2019) 162001.
\newblock \href {http://arxiv.org/abs/1909.11135} {\path{arXiv:1909.11135}},
  \href {https://doi.org/10.1103/PhysRevLett.123.162001}
  {\path{doi:10.1103/PhysRevLett.123.162001}}.

\bibitem{Rathmann:2025jgp}
F.~Rathmann, A.~Nass, K.~O. Eyser, V.~Shmakova, E.~C. Aschenauer, G.~Atoian,
  A.~Cannavo, K.~Hock, H.~Huang, H.~Lovelace, et~al., {Eliminating beam-induced
  depolarizing effects in the hydrogen jet target for high-precision proton
  beam polarimetry at the electron-ion collider}, Phys. Rev. Accel. Beams
  29~(2) (2026) 021001.
\newblock \href {http://arxiv.org/abs/2508.01366} {\path{arXiv:2508.01366}},
  \href {https://doi.org/10.1103/8nh5-l63q} {\path{doi:10.1103/8nh5-l63q}}.

\bibitem{Poblaguev:2020duy}
A.~Poblaguev, A.~Zelenski, G.~Atoian, {The prospects on the absolute proton
  beam polarimetry at EIC}, PoS PSTP2019 (2020) 007.
\newblock \href {https://doi.org/10.22323/1.379.0007}
  {\path{doi:10.22323/1.379.0007}}.

\bibitem{Poblaguev:2024yhl}
A.~Poblaguev, {The Polarized Hydrogen Gas Jet Target. From RHIC to EIC.}, PoS
  SPIN2023 (2024) 091.
\newblock \href {https://doi.org/10.22323/1.456.0091}
  {\path{doi:10.22323/1.456.0091}}.

\bibitem{HERMES:1998twm}
K.~Ackerstaff, A.~Airapetian, N.~Akopov, M.~Amarian, E.~C. Aschenauer,
  H.~Avakian, R.~Avakian, A.~Avetissian, B.~Bains, C.~Baumgarten, et~al., {Beam
  induced nuclear depolarization in a gaseous polarized hydrogen target}, Phys.
  Rev. Lett. 82 (1999) 1164--1168.
\newblock \href {http://arxiv.org/abs/hep-ex/9806006}
  {\path{arXiv:hep-ex/9806006}}, \href
  {https://doi.org/10.1103/PhysRevLett.82.1164}
  {\path{doi:10.1103/PhysRevLett.82.1164}}.

\bibitem{Feynman2011}
R.~P. Feynman, R.~B. Leighton, M.~Sands,
  \href{https://www.feynmanlectures.caltech.edu/}{The Feynman Lectures on
  Physics, Vol. 3}, new millennium Edition, Basic Books, 2011.
\newline\urlprefix\url{https://www.feynmanlectures.caltech.edu/}

\bibitem{Mohr:2024kco}
P.~J. Mohr, D.~B. Newell, B.~N. Taylor, E.~Tiesinga, {CODATA recommended values
  of the fundamental physical constants: 2022*}, Rev. Mod. Phys. 97~(2) (2025)
  025002.
\newblock \href {http://arxiv.org/abs/2409.03787} {\path{arXiv:2409.03787}},
  \href {https://doi.org/10.1103/RevModPhys.97.025002}
  {\path{doi:10.1103/RevModPhys.97.025002}}.

\bibitem{Diermaier:2016fsy}
M.~Diermaier, C.~B. Jepsen, B.~Kolbinger, C.~Malbrunot, O.~Massiczek,
  C.~Sauerzopf, M.~C. Simon, J.~Zmeskal, E.~Widmann, {In-beam measurement of
  the hydrogen hyperfine splitting and prospects for antihydrogen
  spectroscopy}, Nature Commun. 8 (2017) 5749.
\newblock \href {http://arxiv.org/abs/1610.06392} {\path{arXiv:1610.06392}},
  \href {https://doi.org/10.1038/ncomms15749} {\path{doi:10.1038/ncomms15749}}.

\bibitem{Haeberli:1967zr}
W.~Haeberli, {Sources of polarized ions}, Ann. Rev. Nucl. Part. Sci. 17 (1967)
  373--426.
\newblock \href {https://doi.org/10.1146/annurev.ns.17.120167.002105}
  {\path{doi:10.1146/annurev.ns.17.120167.002105}}.

\bibitem{LandauQM}
L.~D. Landau, E.~M. Lifshitz, Quantum Mechanics: Non-Relativistic Theory, 3rd
  Edition, Vol.~3 of Course of Theoretical Physics, Pergamon Press, Oxford,
  1977.

\bibitem{Rabi:1937dgo}
I.~I. Rabi, {Space Quantization in a Gyrating Magnetic Field}, Phys. Rev.
  51~(8) (1937) 652.
\newblock \href {https://doi.org/10.1103/PhysRev.51.652}
  {\path{doi:10.1103/PhysRev.51.652}}.

\bibitem{Wise:2006xj}
T.~Wise, M.~A. Chapman, W.~Haeberli, H.~Kolster, P.~A. Quin, {An optimization
  study for the RHIC polarized jet target}, Nucl. Instrum. Meth. A 556 (2006)
  1--12.
\newblock \href {https://doi.org/10.1016/j.nima.2005.09.042}
  {\path{doi:10.1016/j.nima.2005.09.042}}.

\bibitem{HERMESTargetGroup:2001qci}
C.~Baumgarten, B.~Braun, G.~Court, G.~Ciullo, P.~Ferretti, A.~Golendukhin,
  G.~Graw, W.~Haeberli, M.~Henoch, R.~Hertenberger, et~al., {An atomic beam
  polarimeter to measure the nuclear polarization in the HERMES gaseous
  polarized hydrogen and deuterium target}, Nucl. Instrum. Meth. A 482 (2002)
  606--618.
\newblock \href {https://doi.org/10.1016/S0168-9002(01)01738-7}
  {\path{doi:10.1016/S0168-9002(01)01738-7}}.

\bibitem{Mikirtychyants:2012nh}
M.~Mikirtychyants, R.~Engels, K.~Grigoryev, H.~K. c, P.~Kravtsov, S.~Lorenz,
  M.~Nekipelov, V.~Nelyubin, F.~Rathmann, J.~Sarkadi, et~al., {The Polarized H
  and D Atomic Beam Source for ANKE at COSY-J{\"u}lich}, Nucl. Instrum. Meth. A
  721 (2013) 83--98.
\newblock \href {http://arxiv.org/abs/1212.1840} {\path{arXiv:1212.1840}},
  \href {https://doi.org/10.1016/j.nima.2013.03.043}
  {\path{doi:10.1016/j.nima.2013.03.043}}.

\bibitem{Wise:2004uy}
T.~Wise, M.~Chapman, W.~Haeberli, D.~Graham, A.~Kponou, G.~Mahler, Y.~Makdisi,
  W.~Meng, A.~Nass, J.~Ritter, A.~Zelenski, S.~Kokhanovski, V.~Zubets,
  {Polarized hydrogen jet target for measurement of RHIC proton beam
  polarization}, in: {16th International Spin Physics Symposium (SPIN 2004)},
  2004, pp. 757--760.

\bibitem{ZelenskiPrivate2022}
A.~Zelenski, private communication (2022).

\bibitem{Beijers:2005}
J.~P.~M. Beijers, Adiabatic spin transitions in polarized proton sources,
  Nuclear Instruments and Methods in Physics Research Section A: Accelerators,
  Spectrometers, Detectors and Associated Equipment 536 (2005) 282--288.
\newblock \href {https://doi.org/10.1016/j.nima.2004.08.099}
  {\path{doi:10.1016/j.nima.2004.08.099}}.

\bibitem{LHCspin:2025lvj}
A.~Accardi, A.~Bacchetta, L.~Barion, G.~Bedeschi, V.~Benesova, S.~Bertelli,
  V.~Bertone, C.~Bissolotti, M.~Boglione, G.~Bozzi, et~al., {LHCspin: a
  Polarized Gas Target for LHC} (2025).
\newblock \href {http://arxiv.org/abs/2504.16034} {\path{arXiv:2504.16034}}.

\bibitem{Steffens:2018gzz}
E.~Steffens, \href{https://cds.cern.ch/record/2632904}{{Beam-Induced
  Depolarization and Application to a Polarized Gas Target in the LHC beam}},
  Tech. Rep. CERN-PBC-Note-2018-001, CERN, Geneva (Jul 2018).
\newline\urlprefix\url{https://cds.cern.ch/record/2632904}

\bibitem{Lenisa:2020bxj}
P.~Lenisa, E.~Steffens, V.~Carassiti, G.~Ciullo, P.~Di~Nezza, L.~Pappalardo,
  A.~A. Vasilyev, {LHCspin: a polarized internal target for the LHC}, PoS
  PSTP2019 (2020) 025.
\newblock \href {https://doi.org/10.22323/1.379.0025}
  {\path{doi:10.22323/1.379.0025}}.

\bibitem{Schieck:2008}
H.~{Paetz gen. Schieck}, {Weak field radio-frequency transitions in H and D
  revisited}, Nuclear Instruments and Methods in Physics Research Section A:
  Accelerators, Spectrometers, Detectors and Associated Equipment 587~(2)
  (2008) 213--220.
\newblock \href {https://doi.org/https://doi.org/10.1016/j.nima.2008.01.063}
  {\path{doi:https://doi.org/10.1016/j.nima.2008.01.063}}.

\bibitem{Carroll:1986}
C.~E. Carroll, F.~T. Hioe, {Generalisation of the Landau-Zener calculation to
  three levels}, Journal of Physics A: Mathematical and General 19~(7) (1986)
  1151.
\newblock \href {https://doi.org/10.1088/0305-4470/19/7/017}
  {\path{doi:10.1088/0305-4470/19/7/017}}.

\bibitem{Landau:1932vnv}
L.~D. Landau, {A theory of energy transfer II}, Phys. Z. Sowjetunion 2 (1932)
  46--51.
\newblock \href {https://doi.org/10.1016/B978-0-08-010586-4.50014-6}
  {\path{doi:10.1016/B978-0-08-010586-4.50014-6}}.

\bibitem{Zener:1932ws}
C.~Zener, {Nonadiabatic crossing of energy levels}, Proc. Roy. Soc. Lond. A 137
  (1932) 696--702.
\newblock \href {https://doi.org/10.1098/rspa.1932.0165}
  {\path{doi:10.1098/rspa.1932.0165}}.

\end{thebibliography}

\end{document}